\documentclass[namedreferences]{solarphysics}
\usepackage[hyperref,optionalrh,solaromanenum]{spr-sola-addons} 
\usepackage{graphicx}                   
\usepackage{tabularx}
\usepackage{rotating}
\usepackage[flushleft]{threeparttable}
\usepackage{sidecap}
\usepackage{color}                       
\usepackage{breakurl}                    

\newcommand{\etal}{{\it et al.}}
\newcommand{\eg}{{\it e.g.}}

\newcommand{\aap}{    {\it Astron. Astrophys.}}

\newcommand{\ag}{     {\it Ann. Geophys.}}

\newcommand{\apj}{    {\it Astrophys. J.}}

\newcommand{\apjl}{   {\it Astrophys. J. Lett.}}

\newcommand{\jgr}{    {\it J. Geophys. Res.}}
\newcommand{\mnras}{  {\it Mon. Not. Roy. Astron. Soc.}}

\newcommand{\solphys}{{\it Solar Phys.}}

\newcommand{\ssr}{    {\it Space Sci. Rev.}}

\begin{document}

\begin{article}
\begin{opening}

\title{Variation of Coronal Activity from the Minimum to Maximum of Solar Cycle 24 using Three Dimensional Coronal Electron Density Reconstructions from STEREO/COR1}

\author[addressref={aff1,aff2},corref,email={tongjiang.wang@nasa.gov}]{\inits{T.}\fnm{Tongjiang}~\lnm{Wang}  \orcid{0000-0003-0053-1146}} 
\author[addressref={aff1,aff2},corref,email={nelson.l.reginald@nasa.gov}]{\inits{N.L.}\fnm{Nelson L.}~\lnm{Reginald}}
\author[addressref={aff2},corref]{\inits{J.M.}\fnm{Joseph M.}~\lnm{Davila}}
\author[addressref={aff2},corref]{\inits{O.C.}\fnm{O. Chris}~\lnm{St. Cyr}}
\author[addressref={aff3},corref]{\inits{W.T.}\fnm{William T.}~\lnm{Thompson}}
\address[id=aff1]{Department of Physics, Catholic University of America,
   620 Michigan Avenue NE, Washington, DC 20064, USA}
\address[id=aff2]{NASA Goddard Space Flight Center, Code 671, Greenbelt, MD 20771, USA}
\address[id=aff3]{ADNET Systems, Inc., NASA Goddard Space Flight Center, Code 671, Greenbelt, MD 20771, USA}
%
 \runningauthor{T. Wang \etal}
 \runningtitle{Coronal Activity from the Minimum to Maximum of Solar Cycle 24}

\begin{abstract}

 Three dimensional electron density distributions in the solar corona are reconstructed for 100 Carrington Rotations (CR 2054--2153) during 2007/03--2014/08 using the spherically symmetric method from polarized white-light observations with the {\it inner coronagraph} (COR1) onboard the twin {\it Solar Terrestrial Relations Observatory} (STEREO). These three-dimensional electron density distributions are validated by comparison with similar density models derived using other methods such as tomography and a magnetohydrodynamic (MHD) model as well as using data from {\it Solar and Heliospheric Observatory} (SOHO)/{\it Large Angle and Spectrometric Coronagraph} (LASCO)-C2. Uncertainties in the estimated total mass of the global corona are analyzed based on differences between the density distributions for COR1-A and -B. Long-term variations of coronal activity in terms of the global and hemispheric average electron densities (equivalent to the total coronal mass) reveal a hemispheric asymmetry during the rising phase of Solar Cycle 24, with the northern hemisphere leading the southern hemisphere by a phase shift of 7--9 months. Using 14-CR ($\approx$13-month) running averages, the amplitudes of the variation in average electron density between Cycle 24 maximum and Cycle 23/24 minimum (called the modulation factors) are found to be in the range of 1.6--4.3. These modulation factors are latitudinally dependent, being largest in polar regions and smallest in the equatorial region. These modulation factors also show a hemispheric asymmetry, being somewhat larger in the southern hemisphere. The wavelet analysis shows that the short-term quasi-periodic oscillations during the rising and maximum phases of Cycle 24 have a dominant period of 7--8 months. In addition, it is found that the radial distribution of mean electron density for streamers at Cycle 24 maximum is only slightly larger (by $\approx$30\%) than at cycle minimum.

\end{abstract}

%
\keywords{Solar corona $\cdot$ electron density $\cdot$ solar cycle $\cdot$ oscillations $\cdot$ STEREO $\cdot$ COR1}
\end{opening}

%
\section{Introduction}

The solar cycle is the long-term ($\approx$11-year) variation of solar activity, manifested in various phenomena such as sunspot numbers, flares, coronal mass ejections (CMEs), and total solar radiation  (see \inlinecite{hat15} for a review). It is a well accepted fact that the solar cycle is virtually the magnetic cycle (with the full period of $\approx$22 years due to the 11-year polarity reversal of sunspots and polar fields) and is produced by dynamo processes within the Sun (see \inlinecite{cha10} for a review). At solar minimum a dipolar field dominates the large-scale structure of the solar corona, and is characterized by a long-lived helmet streamer belt and large polar coronal holes (CHs), whereas during solar maximum higher order components of the magnetic field strengthen, resulting in large-scale coronal structures that are increased in complexity \citep{lin99, ril06, hu08, yea13}. This is indicated by a widening of the streamer belt to higher latitudes and emergence of pseudo- and polar streamers. The solar magnetic field is understood to play a crucial role in forming the structure of the solar corona and inner heliosphere (see \inlinecite{lin99} and \inlinecite{mac12} for a review). However, direct measurements of the weak coronal magnetic field are still impossible. To obtain the global magnetic structure of the solar corona we mainly rely on the simple potential field source surface (PFSS) model \citep{sch69, alt69, sch03} and more advanced magnetohydrodynamic (MHD) models \citep{ril06, hu08, lio09, rus10}. To validate these models, the calculated coronal magnetic structures are often compared with white-light observations of particular coronal brightness structures such as helmet streamers and coronal holes. This is because the coronal plasma outlines the direction of the magnetic field in the highly conducting solar corona where the plasma and magnetic field are effectively frozen together. As a result, systematic long-term observations of coronal brightness structure should be able to reflect morphological/topological changes of the large-scale coronal magnetic field over the solar cycle. For example, using a three-dimensional (3D) MHD model with  the observed line-of-sight (LOS) photospheric magnetic field as boundary conditions, \citet{hu08} studied the evolution of the heliospheric current sheet (HCS) and the coronal magnetic field configuration during Cycle 23, and confirmed the close spatial relationship between the observed white-light streamer structures and the HCS (\eg~ \inlinecite{guh96} and references therein). 

The K-coronal brightness variation (and by inference, the variation of the coronal mass) with solar cycle was first detected from eclipse observations, and was followed with systematic studies using long-term observations by ground- and space-based coronagraphs. For example, from inner coronal observations (at 1.3 and 1.5 R$_\odot$ from the Sun center), \citet{fis84} deduced that the integrated polarized brightness (pB) of the K corona increased by a factor of 2 from the minimum to maximum during Cycles 20 and 21. They also found that the total coronal hole area increased from zero at solar maximum to a value up to about one quarter of the global area at solar minimum. Using the outer coronal observations (at 2.0--3.4 R$_\odot$) from the {\it Solar Maximum Mission} (SMM) coronagraph during Solar Cycle 22, \citet{mac01} showed that the K-coronal brightness (or mass) varied by a factor of 4 between the solar maximum and minimum, and was closely correlated with the occurrence rate and average mass of CMEs. Using white-light observations from the {\it Large Angle and Spectrometric Coronagraph} (LASCO)-C2 on the {\it Solar and Heliospheric Observatory} (SOHO) spacecraft, \citet{lam14} compared the solar activity minima of Cycles 22/23 and 23/24, and \citet{bar15} studied the variability of the K-coronal radiance over timescales ranging from mid-term (0.6--4 years) quasi-periodicities to the long-term solar cycle. The mid-term quasi-periodicities appear to be a basic property of the solar activity. This was suggested by the fact that many features are common to different observations in the solar atmosphere and even in the convective zone (see \inlinecite{baz14} for a review). In addition, the observed hemispheric asymmetry is another important property of the solar activity to be understood \citep{uso09, mci13, baz14, ric16, nor14}. 

\begin{SCfigure} 
\centering
\includegraphics[width=0.5\textwidth]{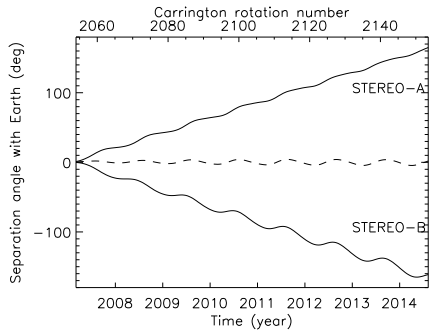}
\caption{Separation angles of the STEREO-A/B spacecraft with the Earth during the period from 2007/03/04 to 2014/08/07. The positive and negative values indicate STEREO-A and -B, respectively. The dashed line indicates the median position of the separation angle between the two spacecraft relative to Earth. } 
\label{fig:stpos}
\end{SCfigure}

In this paper, we present the variations of coronal activity during Solar Cycle 24 based on the 3D electron density distribution derived from pB observations of the K-corona using {\it Sun Earth Connection Coronal and Heliospheric Investigation} (SECCHI)-{\it inner coronagraph} (COR1) telescopes \citep{tho03, tho08} onboard the twin {\it Solar Terrestrial Relations Observatory} (STEREO) spacecraft \citep{how08}. In Section~\ref{sctden} we describe the method used to reconstruct the 3D electron density of the corona, and in Section~\ref{sctval} we validate the reconstructed density models using several techniques. In Section~\ref{sctana} we then analyze the hemispheric asymmetry and the short-period oscillations of the average electron density (or coronal mass) during the rising and maximum phases of Cycle 24, and the associated variations in the streamer area and electron density. In Section~\ref{sctdc} we present discussion and conclusions from our study.    

\section{3D Electron Density Reconstructions Using STEREO/COR1}
\label{sctden}
We reconstruct 3D distributions of the coronal electron density for 100 Carrington Rotations (CRs) from CR 2054 to CR 2153 using the pB data acquired by STEREO/COR1. During this period both STEREO spacecraft ran in the heliocentric orbit that is close to the Earth's. Spacecraft-A (Ahead) drifts away from Earth in the direction of the Earth's rotation with an orbital period slightly shorter than a year, while spacecraft-B (Behind) drifts away from Earth in the opposite direction with an orbital period slightly longer than a year. The two spacecraft separate from each other at an average rate of approximately 45$^{\circ}$ per year. Figure~\ref{fig:stpos} shows that STEREO-A and STEREO-B were separated from the Earth by between 1.2$^{\circ}$--165.9$^{\circ}$ and 0.2$^{\circ}$--160.9$^{\circ}$, respectively, during the period of interest, {\it i.e.} from 04:14 UT on 4 March 2007 (the beginning time of CR 2054 viewed from Earth) to 11:05 UT on 21 August 2014 (the ending time of CR 2153). COR1 observes the white-light K corona from about 1.4 to 4 R$_\odot$ in a waveband of 22.5 nm wide centered on the H$\alpha$ line at 656 nm. The data are taken with a cadence of 10 minutes and transmitted in the binning format of 1024$\times$1024 from the spacecraft. From 19 April 2009 the normal cadence is increased to 5 minutes and the binned images are in the 512$\times$512 format. The instrumental scattered light in the pB data is removed by subtracting the combined monthly minimum and calibration roll backgrounds \citep{tho10}. 

We use the spherically symmetric polynomial approximation (SSPA) method to reconstruct the 3D coronal density. The SSPA method is based on the assumption that the radial electron density distribution has the polynomial form, $N(r)=\sum_k a_k r^{-k}$, where $r$ is the radial distance from the Sun center, $k$ is the degree of the polynomial, and $a_k$ are unknown coefficients \citep{hay01, wan14}. The coefficients $a_k$ can be determined by a multivariate least-squares fit to the radial profile of pB data. \citet{wan14} validated the SSPA method using synthesized pB images from a 3D density model reconstructed by tomography from COR1 observations, and showed that the derived density is consistent with model inputs in the plane of the sky (POS) generally within a factor of two. The degree of $k=5$ is typically suitable for COR1 pB inversion. In addition, \citet{wan14} also demonstrated a reconstruction of the 3D coronal density using the SSPA method. Here we adopt a similar procedure as described below.

Each reconstruction is made of 2D density maps ($N(r,\theta)$) with the radial and latitudinal dependence in the POS. These 2D density maps are inverted by the SSPA method from a set of pB data (typically including 56 images with a cadence of 6 hours that corresponds to a longitudinal step of about 3$^\circ$) over a period of 13.6 days. Thus, two reconstructions are made for a given CR. If an image at some sampled time is missing or bad, it is replaced with the one observed closest to that time. First, a 2D density map is derived by fitting the radial pB data between 1.6 and 3.7 R$_\odot$ using the SSPA inversion at 120 angular positions (with intervals of 3$^\circ$) surrounding the Sun for each image. Then the east-limb and west-limb density profiles of all the images are mapped into spherical cross sections at different heights (with an interval of 0.1 R$_\odot$) based on their Carrington coordinates. After converting the irregular grid into the regular grid ($2^\circ\times2^\circ$ in longitude and latitude), a 3D density reconstruction in the radial range of 1.5--3.7 R$_\odot$ is obtained finally.    

\begin{figure}
  \centerline{\includegraphics[width=1.\textwidth,clip=]{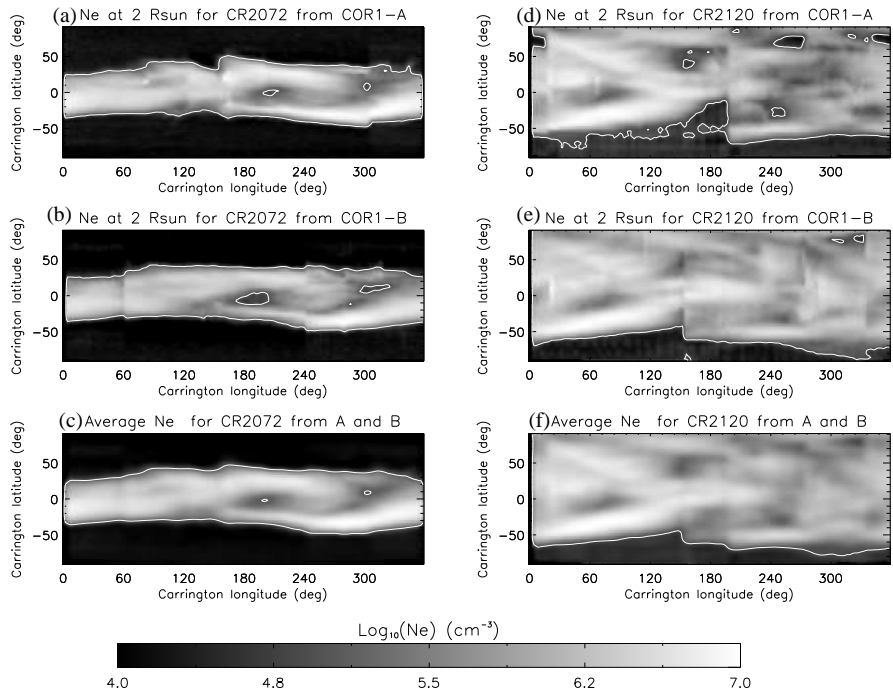}}
  \caption{ Two examples of the 3D coronal electron density reconstructed by the SSPA method. Left Panels: for CR 2072 during solar minimum, showing a spherical cross section of the density at 2.0 R$_\odot$ from COR1-A (top), COR1-B (middle), and the mean of COR1-A and -B with $10^{\circ}\times10^{\circ}$ smoothing (bottom). Right Panels: Same as the left panels but for CR 2120 during solar maximum. The overlaid contour encloses regions with density $N_{\rm e}>3\sigma$ (see Figure~\ref{fig:nsgm}). An animation of 200 reconstructions for CRs 2054--2153 during 2007-2014 is available in the online version of the journal. }
\label{fig:nemap}
\end{figure}

By applying this method to STEREO/COR1 pB observations during 2007--2014, we obtain 200 reconstructions of the 3D coronal density for CRs 2054--2153 from COR1-A and COR1-B, respectively. We average the reconstructions for COR1-A and -B and construct a $10^\circ\times10^\circ$ smoothed coronal density map to compare with coronal density models determined by other methods (see Section~\ref{sctval}). Figure~\ref{fig:nemap} shows two examples, one for CR 2072 for the period of 20 July--3 August 2008 during the solar minimum, the other for CR 2120 for the period of 6--19 February 2012 during the solar maximum. An animation showing all density reconstructions at 2 R$_\odot$ for CRs 2054--2153 from COR1-A and -B as well as their average is available in the online version of the journal.

\section{Validations}
\label{sctval}
In the following sections we compare our reconstructed 3D distributions of coronal electron density with several different techniques. These include the derivation based on LASCO C2, tomographic reconstruction, and MHD modeling. We also analyze the sources of uncertainty in our estimated total coronal mass for these reconstructions.

\subsection{LASCO/C2 pB Inversion}
SOHO/LASCO-C2 has typically made one pB sequence per day since late 1995 \citep{bru95}. The C2 has an effective field of view (FOV) of 2.2--6.1 R$_\odot$ \citep{fra12}, and overlaps with that of STEREO/COR1. This allows us to use the 2D coronal density $N(r,\theta)$ inverted from the C2 pB images to validate the reconstructed 3D coronal density from COR1. Since the COR1's 3D density is essentially made using the sequence of 2D density maps, comparisons with the C2's 2D density cannot provide direct examination of 3D characteristics of the corona, but may allow us to test the requirement that main coronal structures need to be stable over about two weeks for reconstruction. This is because the 3D density reconstructions presented here are the average between COR1-A and COR1-B, and the 2D density distributions used for comparison between COR1-A/B and C2 in the same POS ({\it i.e.} when viewed from the same direction in Carrington coordinate system) are observed at different times (see the following examples).

\begin{figure}
\centerline{\includegraphics[width=0.9\textwidth,clip=]{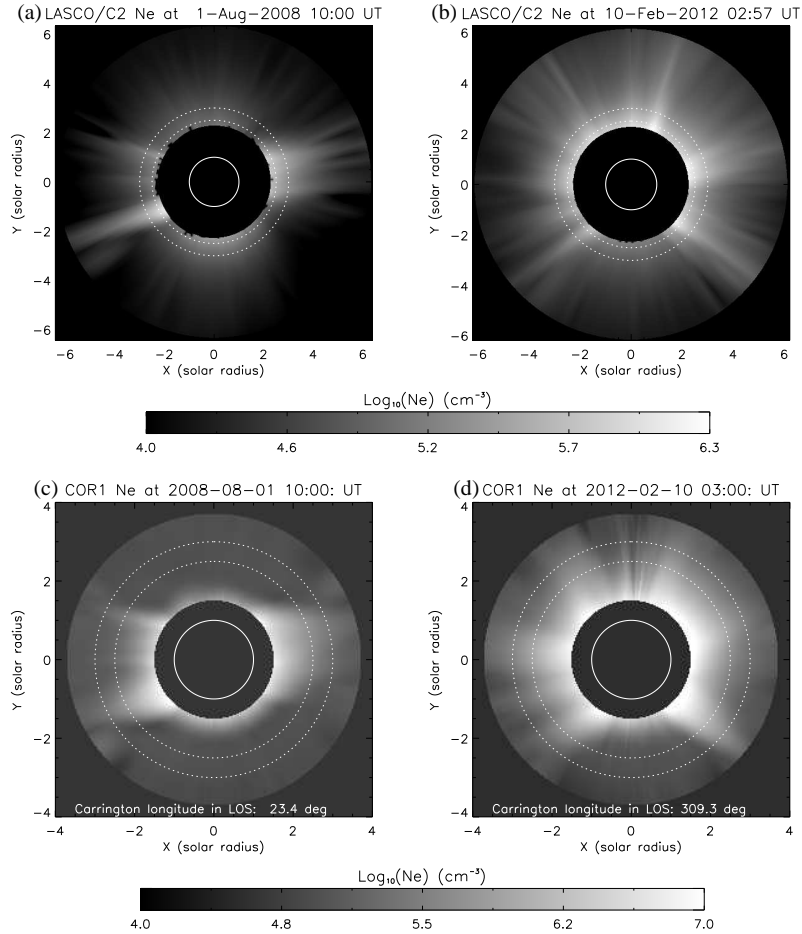}}
\caption{Comparison between the STEREO/COR1 and LASCO/C2 coronal densities.
(a) and (b): 2D electron density maps derived by Van de Hulst (VdH) inversion from LASCO/C2 pB images observed at 10:00 UT on 1 August 2008 and at 02:57 UT on 10 February 2012, respectively. (c) and (d): cross sections of the STEREO/COR1 3D electron density at POS, corresponding to positions of the Earth at the LASCO observing times in (a) and (b), respectively. In (c) and (d) the 3D coronal density for CR 2072 and CR 2120 is used, respectively, and the Carrington longitude of the LOS direction (23.4$^{\circ}$ in (c) and 309.3$^{\circ}$ in (d)) is marked at the bottom. In each panel the solid circle indicates the solar limb and the two dotted circles (at 2.5 and 3.0 R$_\odot$) mark the paths along which the density profiles are shown in Figure~\ref{fig:nlpa}.} 
\label{fig:nlsc}
\end{figure}

\begin{figure}
\centerline{\includegraphics[width=1.\textwidth,clip=]{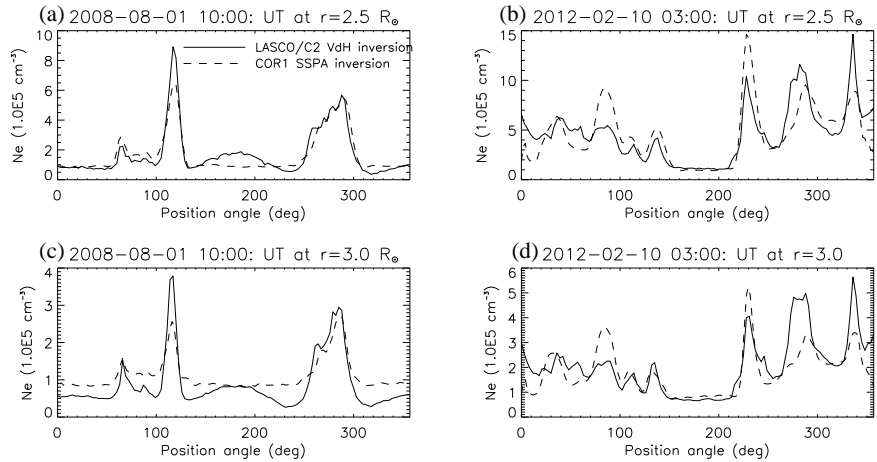}}
\caption{Comparison between the electron densities along two circular paths at 2.5 R$_\odot$ (a)--(b), and 3.0 R$_\odot$ (c)--(d) from the LASCO/C2 and STEREO/COR1 2D density maps. The position angle is counted anticlockwise from the north pole. Panels (a) and (c) correspond to the case where LASCO/C2 data are observed at 10:00 UT on 1 August 2008, and panels (b) and (d) where LASCO/C2 data are observed at about 03:00 UT on 10 February 2012. In each panel the solid line represents the densities from the LASCO/C2 VdH inversion, and the dashed line represents the densities from the STEREO/COR1 SSPA inversion.} 
\label{fig:nlpa}
\end{figure}

We used the calibrated C2 pB images which are available on the NRL archive\footnote{http://lasco-www.nrl.navy.mil/content/retrieve/polarize/}. We chose the 3D density reconstructions for CR 2072 and CR 2120 as examples (see Figure~\ref{fig:nemap}). We use the routine {\scshape pb\_inverter} in SolarSoftWare (SSW; see \inlinecite{fre98}) to derive the 2D coronal density distribution from the C2 pB images. This routine uses the Van de Hulst (VdH) technique \citep{van50}. In the VdH inversion the radial distribution of pB in the POS is assumed to follow a polynomial function, while in the SSPA inversion the electron density distribution is assumed to follow a polynomial function. \citet{wan14} showed in theory and observation that these two methods are equivalent. We modify the code {\scshape pb\_inverter} by replacing the use of the IDL function {\scshape curvefit} with {\scshape svdfit} in fitting the radial pB data to a polynomial function with the degree $k$ equal to four, because the {\scshape svdfit} works better in the case of computing a linear least squares fit. Figures~\ref{fig:nlsc}a and~\ref{fig:nlsc}b show the density maps derived from the C2 pB images observed at 10:00 UT on 1 August 2008 during solar minimum and at about 03:00 UT on 10 February 2012 during solar maximum, respectively. For comparison we make the 2D density maps from the COR1 3D density model by calculating its cross sectional distribution at the POS as viewed from Earth at the observing time for LASCO/C2 images. Figure~\ref{fig:nlsc}c shows a density map from CR 2072 that is equivalent to the average between $N(r,\theta)$ obtained from COR1-A at 2008/07/21 09:00 UT and that of COR1-B at 2008/07/30 06:00 UT. Figure~\ref{fig:nlsc}d shows, in the case of CR 2120, an equivalent average between the density maps from COR1-A at 2012/02/18 9:00 UT and from COR1-B at 2012/02/15 00:00 UT. The 2D density distributions from COR1 and LASCO/C2 are found to be consistent. For quantitative comparison, we plot the COR1 and C2 density profiles as a function of position angles at two heights (2.5 and 3.0 R$_\odot$) in Figure~\ref{fig:nlpa}. The comparison indicates a good coincidence in position and width between streamers in the COR1 and C2 density maps for both solar minimum and maximum cases. The differences in the peak density are within a factor of two, comparable to the uncertainty from the SSPA inversion process \citep{wan14}.

\begin{figure}
\centerline{\includegraphics[width=1.\textwidth,clip=]{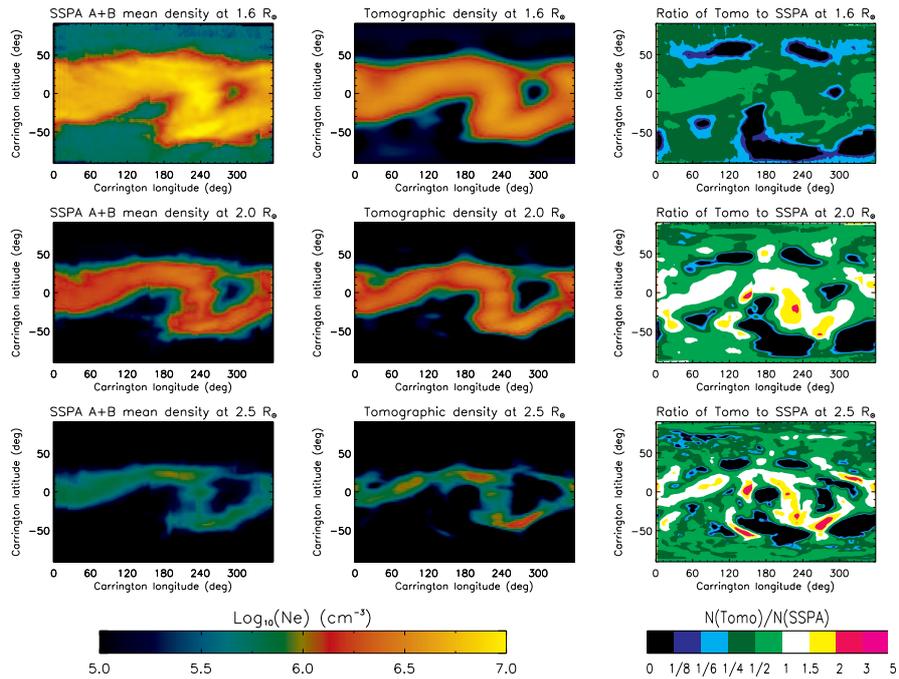}}
\caption{Comparison between the 3D coronal electron densities for CR 2066 reconstructed by SSPA and tomography methods. Left Panels: spherical cross sections of the SSPA mean density for COR1-A and -B at 1.6, 2.0, and 2.5 R$_\odot$ (from top to bottom).  Middle Panels: Same as the left panels but for the density reconstructed by tomography from COR1-B. Right Panels: The ratio of the tomographic density (2nd column) to SSPA density (1st column). }
\label{fig:ntom}
\end{figure}

\subsection{Validation with Tomography}
The tomographic technique reconstructs optically thin 3D coronal density structures using observations from multiple viewing directions, or using observations gathered over a period of half a solar rotation by a single spacecraft or only from Earth-based coronagraphs \citep[\eg][]{fra02, fra07, fra10, kra09, bar13,vib16}. Generally, only structures stable over about two weeks can be reliably reconstructed from tomographic techniques. \citet{kra09} reconstructed a 3D coronal electron density model for CR 2066 for the period of 1--14 February 2008 based on 28 pB images (with a cadence of about 2 images per day) from COR1-B using the regularized tomographic inversion method. \citet{wan14} compared this tomographic reconstruction with the SSPA reconstruction based on the same dataset, and found them to be consistent. The ratios of the tomographic density to the SSPA density in the streamer belt are very close to 1, typically in the range 1/2--2. Here we reconstruct the SSPA 3D coronal density for the same period but consisting of 55 pB images (about 4 images per day) from COR1-A and -B, respectively. The mean density distributions for COR1-A and -B show a good agreement with those by tomography obtained with 28 pB images at different heliocentric distances (see Figure~\ref{fig:ntom}). The reason why using pB data with higher cadence does not help to improve the actual spatial resolution of the reconstructed density in longitude is that the SSPA method has an intrinsic limitation (with angular resolution of $\approx50^{\circ}$) in resolving the coronal structure near the POS due to the spherically symmetric approximation in inversion \citep{wan14}. 

\begin{SCfigure}
\centering
\includegraphics[width=0.6\textwidth]{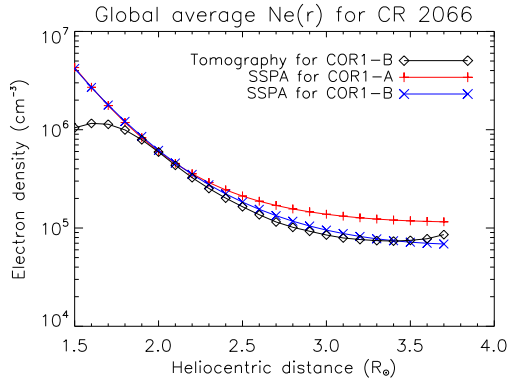}
\caption{Radial profiles of the globally-averaged coronal density for the SSPA and tomographic reconstructions for CR 2066. The values determined by tomography from COR1-B are shown with diamonds, and those by SSPA from COR1-A and COR1-B are shown with pluses and crosses, respectively. }
\label{fig:ntmr}
\end{SCfigure}

In addition, it is particularly useful to compare the globally averaged radial density profiles between the SSPA and tomography reconstructions because it helps to determine whether their coronal mass distributions are consistent overall in the analyzed volume despite the local difference. Figure~\ref{fig:ntmr} shows that the globally-averaged density profile for tomography is consistent with the SSPA for COR1 within 1.8--3.7 R$_\odot$. The better consistency with COR1-B than COR1-A is because the tomographic reconstruction is made from the COR1-B data. It is estimated that the radial density for COR1-A is larger (by a factor of about 1.6) than for COR1-B in the outer part of the FOV (2.7--3.7 R$_\odot$). This difference may be explained by the fact that the COR1-B instrumental background is substantially lower than COR1-A before 30 January 2009. After that date the level of scattering in COR1-A and -B becomes comparable (see Figure~\ref{fig:nmdr}), likely due to contamination of the COR1-B objective by a dust particle (see the discussion in Section~\ref{scterr} and in~\inlinecite{tho10}). In addition, the tomographic reconstruction used here may underestimate the density near the occulter by a factor of about 2--3 due to the boundary effect since the solution at the grid points close to the occulter is less constrained by the observational data.

\begin{figure}
\centerline{\includegraphics[width=1.\textwidth,clip=]{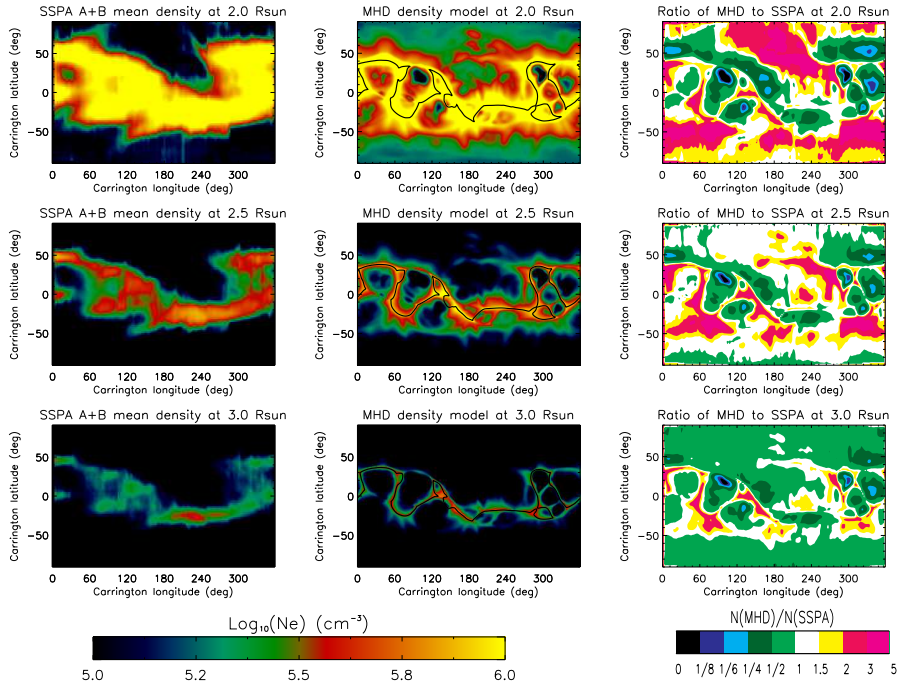}}
\caption{Comparison between the 3D coronal electron densities for CR2097/2098 derived by the SSPA inversion and MHD simulation. Left Panels: spherical cross sections of the SSPA mean density from COR1-A and -B at 2.0, 2.5, and 3.0 R$_\odot$ (from top to bottom).  Middle Panels: Same as the left panels but for the electron density from a MHD simulation of the solar eclipse of 11 July 2010 using SOHO/MDI magnetic field data. The overlaid solid lines denote the magnetic neutral line. Right Panels: The ratio of the MHD simulated density (2nd column) to SSPA density (1st column).} 
\label{fig:nmod}
\end{figure}

\subsection{Validation with the MHD Simulation}
In this section we show an example of validating the SSPA reconstruction with the MHD simulated coronal model. We use the Corona Heliosphere (CORHEL) and Magnetohydrodynamics Around a Sphere (MAS) model (known as the CORHEL MAS model) developed by Predictive Science Inc. (see \inlinecite{mik99} for the details). The CORHEL MAS model is a sophisticated global thermodynamic MHD model that uses an improved equation for energy transport in the corona that includes parameterized coronal heating, parallel thermal conduction along the magnetic field lines, radiative losses, and acceleration by Alfv\'{e}n wave \citep{mik07}. The global plasma density and temperature structures simulated by this model are capable of reproducing major coronal features observed in extreme ultraviolet (EUV) and X-ray emission, and have been successfully used to predict the white-light coronal structures for many total solar eclipses \citep[\eg][]{lio09, rus10}.  
 
The middle panels of Figure~\ref{fig:nmod} show the modeled electron density at heliocentric distances of 2.0, 2.5, and 3.0 R$_\odot$ from a thermodynamic MHD simulation. This simulation is carried out for predicting the coronal structure of the 11 July 2010 eclipse, which used the photospheric magnetic field data measured with SOHO/MDI during a period from 10 June to 4 July 2010 (a combination of CR 2097 and 2098) as boundary conditions. The artificial data produced based on the simulated results were also used to test the tomography method \citep{kra14}, and to estimate uncertainties of the Spherically Symmetric Model (SSM) in determining the electron temperatures and bulk flow speeds in the low corona \citep{reg14}. 

\begin{SCfigure}
\centering
\includegraphics[width=0.6\textwidth]{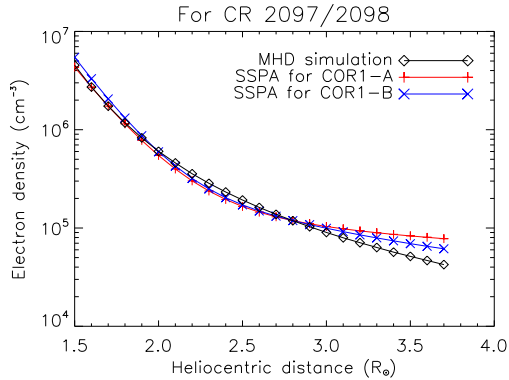}
\caption{Comparison between the radial profiles of globally averaged electron density from the 3D distributions by SSPA and the MHD model. The values from the MHD model are shown with diamonds, and those from the SSPA COR1-A and COR1-B are shown with pluses and crosses, respectively.  }
\label{fig:nmdr}
\end{SCfigure}

To compare with the MHD simulated coronal density, we make two SSPA reconstructions for the period of 9 June--7 July 2010 using 111 pB images from COR1-A and -B, respectively, and then average them to obtain a mean density model. The left panels of Figure~\ref{fig:nmod} show the electron density distributions at 2.0, 2.5, and 3.0 R$_\odot$ from the SSPA density model. We find that the streamer regions with high densities are mainly located along the magnetic neutral lines. The SSPA and simulated density distributions are overall consistent, but the simulated density has more fine structures (see middle panels of Figure~\ref{fig:nmod}). Figure~\ref{fig:nmdr} compares the globally-averaged radial density profiles in the range 1.5--3.7 R$_\odot$. The SSPA densities for both COR1-A and -B are consistent with the simulated results except for the region close to the outer part ($>$3.5 R$_\odot$) of the COR1 FOV where the SSPA density values are a little bit higher. This is probably because the pB data in that region have weak signal-to-noise ratios, leading to an inverted density signal just above the background noise level (see Figure~\ref{fig:nnrv}b in Section~\ref{sctstr}). 

The median position between STEREO-A and -B in heliographic longitude varies around the Earth with amplitudes less than 4$^{\circ}$ (see the dashed curve in Figure~\ref{fig:stpos}). This may account for the reasonable comparison between the SSPA density model obtained by averaging COR1-A and -B reconstructions and the model calculated from the MHD simulation using the SOHO/MDI magnetic field data. 

\begin{figure}
\centerline{\includegraphics[width=1.0\textwidth,clip=]{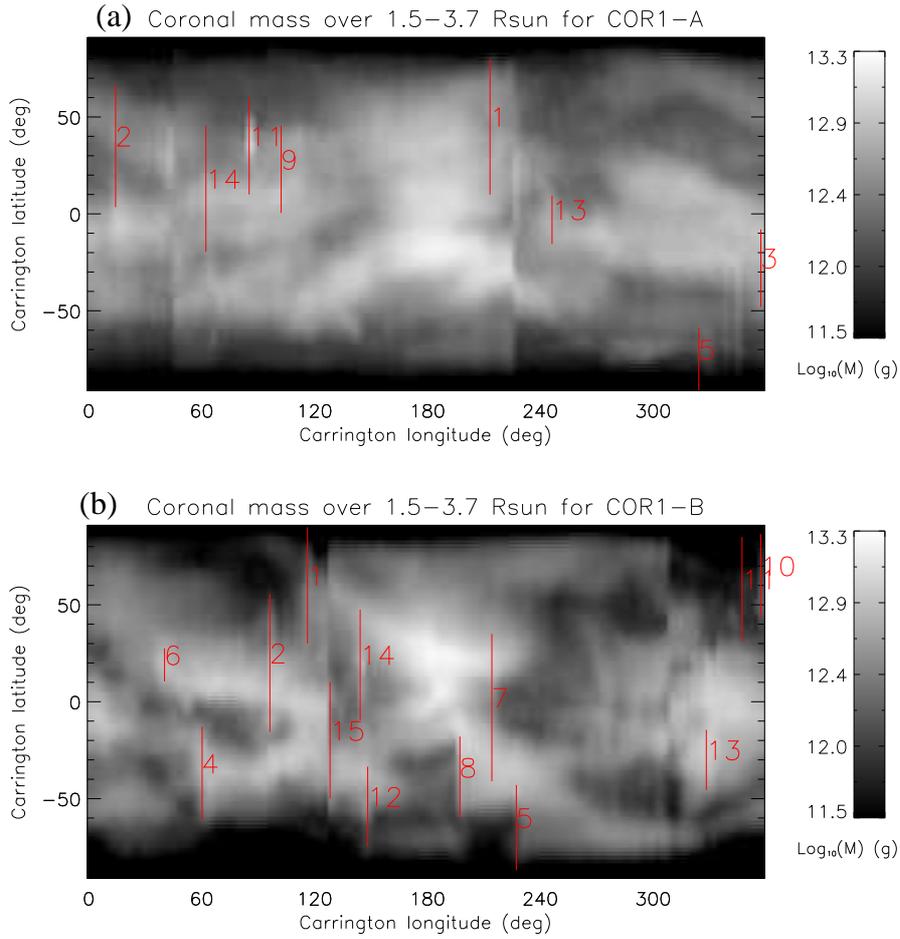}}
\caption{The radially-integrated coronal mass distributions of CR 2136 for the period of 1--15 May 2013 from COR1-A (a) and COR1-B (b). The vertical solid bars indicate the location and latitudinal width of CMEs that are listed in Table~\ref{tab:cme}. An animation for CRs 2054--2153 is available in the online version of the journal.}
\label{fig:cme}
\end{figure}

\subsection{Error Analysis}
\label{scterr}
The reconstructions of the 3D coronal electron density from STEREO/COR1 are subject to several sources of uncertainties and error including i) the effect of CMEs or other transient phenomena (\eg~coronal dimmings); ii) the temporal evolution of coronal structures within a given period; iii) the instrumental background subtraction; and iv) the spherically symmetric approximation in the SSPA inversion.

\begin{figure}
\centerline{\includegraphics[width=0.8\textwidth,clip=]{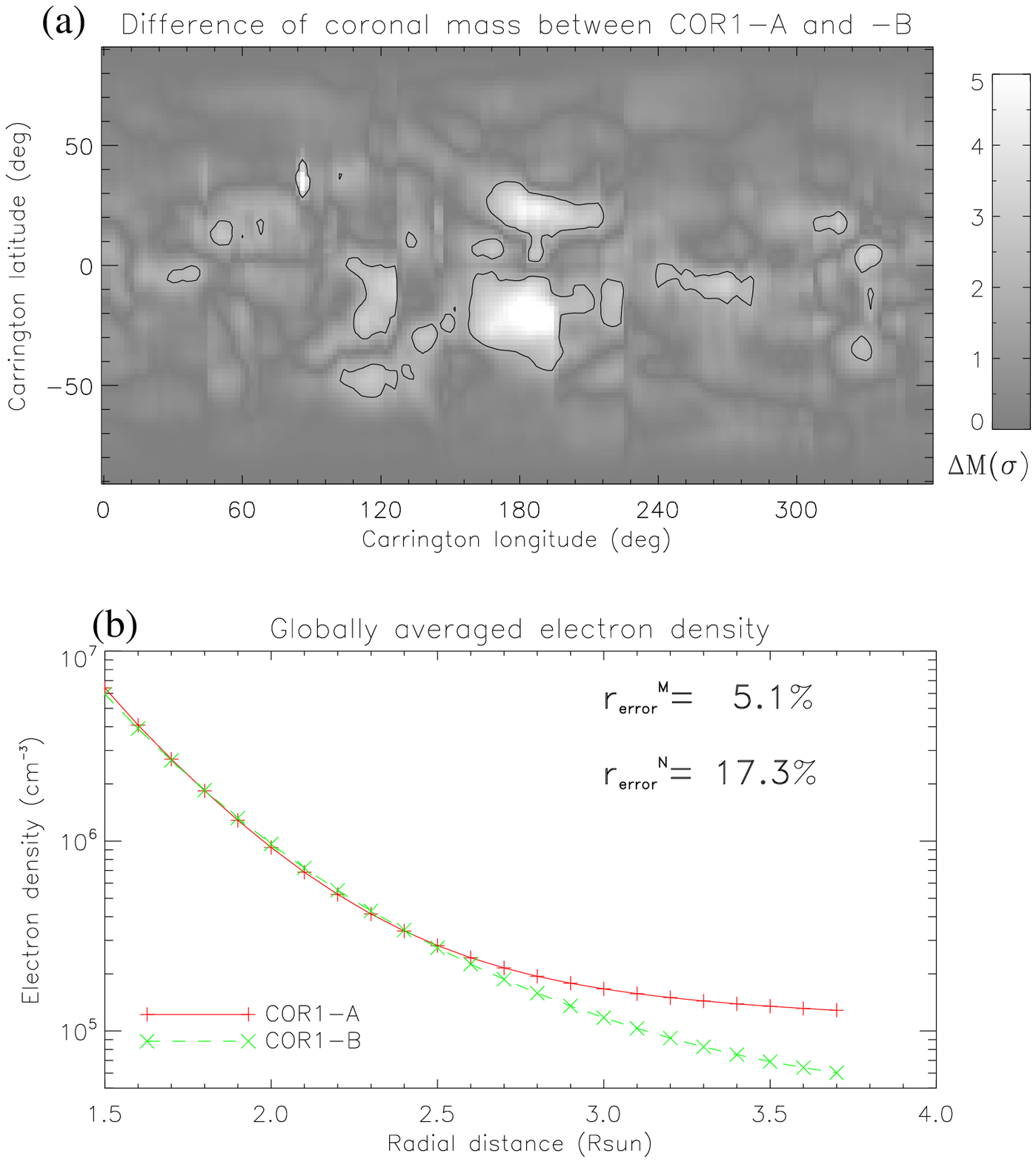}}
\caption{ (a) The mass difference map calculated between COR1-A and -B reconstructions of CR 2136 for the period of 1--15 May 2013. The color bar is in units of the standard error ($\sigma$) for the global average. The overlaid contour encloses regions of the mass difference with $|\Delta{M}(i,j)|>2\sigma$. (b) Comparison of the radial profiles of globally-averaged coronal density for COR1-A and -B. The marked $r_{\rm error}^M$ and $r_{\rm error}^N$ are the errors for the total coronal mass and average density calculated using Equations~(\ref{equ:erm}) and (\ref{equ:ern}), respectively. An animation for CRs 2054--2153 is available in the online version of the journal.}
\label{fig:dif}
\end{figure}

\begin{figure}
\centerline{\includegraphics[width=0.9\textwidth,clip=]{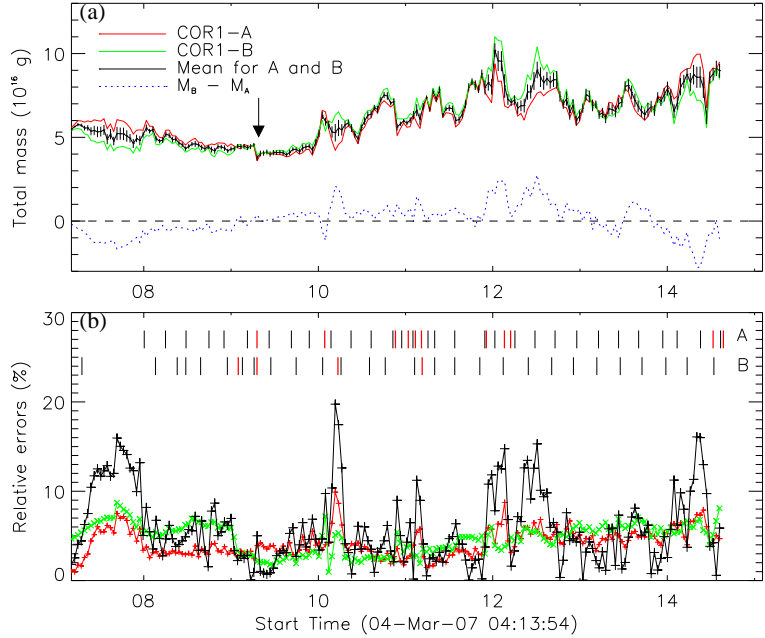}}
\caption{ Estimate of errors in the measured total coronal mass from STEREO/COR1. 
(a) Temporal variations of the total coronal mass integrated from 1.5 to 3.7 R$_\odot$ globally (the red line for COR1-A ($M_A$), the green line for COR1-B ($M_B$), and the black line for their mean). The error bars are calculated using Equation~(\ref{equ:erm}). The dotted curve represents the total mass differences ($M_B-M_A$) between COR1-A and -B. The arrow marks a small drop in the total mass when the binning format of COR1 data is changed from $1024\times1024$ to $512\times512$ on 19 April 2009. (b) The black curve with the $+$ symbols represents the relative mass differences ($|M_B-M_A|/(M_B+M_A)$) between COR1-A ans -B, the red curve with the $+$ symbols represents the errors ($r_{\rm error}^M$) for the total coronal mass calculated using Equation~(\ref{equ:erm}), and the green curve with the $\times$ symbols represents the errors ($r_{\rm error}^N$) for the globally-averaged coronal density calculated using Equation~(\ref{equ:ern}). Note that the curve $r_{\rm error}^N$ is scaled by a factor of 1/3 for better comparison with $r_{\rm error}^M$. The thin black bars at the top of panel (b) indicate the times of  calibration roll maneuvers for COR1-A (upper row) and COR1-B (lower row) listed in Table~\ref{tab:calr}. The thick red bars indicate the times of the events that cause the COR1 background change listed in Table~\ref{tab:evt}.} 
\label{fig:err}
\end{figure}

We first chose CR 2136, a density reconstruction during the maximum period of solar activity, as an example to detail the method of error analysis, then we show the results for the uncertainties for all rotations. Figure~\ref{fig:cme} shows the coronal mass distributions of CR 2136 for COR1-A and -B, which are calculated by integrating the 3D density in the region of 1.5--3.7 R$_\odot$ using
\begin{equation}
M(i,j)=\mu m_p R_{\odot}^3\Delta{\phi}\Delta{\theta}\Delta{r}\sum_{k}N(i,j,k)r_k^2{\rm cos}\theta_j, \label{equ:mss}
\end{equation}
where the electron density is assumed to be equal to the ion density, $m_p$ is the mass of the proton, $\mu$=1.2 is the mean molecular weight in the corona, $N(i,j,k)$ is the electron density value at a grid point ($i,j,k$), $\Delta{\phi}$, $\Delta{\theta}$, $\Delta{r}$ are the grid size in longitude, latitude, and radius, respectively, $r_k$ is the dimensionless radial distance, and $\theta_j$ is the latitude. For a 3D density reconstruction, $\Delta{\phi}=\Delta{\theta}=2^{\circ}(\pi/180^{\circ})$ rad, $\Delta{r}$=0.1, $r_k$=[1.5, 3.7], and $\theta_j$=[$-90^{\circ}$, $90^{\circ}$]. We calculate the total coronal mass for CR 2136 (and other rotations; see Figure~\ref{fig:err}a) by integrating the global corona:
\begin{equation}
M_{\rm total}=\sum_{i,j}M(i,j).
\label{equ:mtt}
\end{equation}

We identify 15 CMEs from the pB images used for the reconstruction of CR 2136 based on the GSFC COR1 CME catalog\footnote{https://cor1.gsfc.nasa.gov/catalog} which are marked in Figure~\ref{fig:cme}. Table~\ref{tab:cme} lists the observing time, Carrington coordinate of the center position ($L_{\rm c}$, $B_{\rm c}$), and latitudinal width ($W$) of these CMEs. Since the 3D density reconstruction is made with the pB images with a cadence of about 6 hours, each CME showed up only in one frame but may cover 2--3 grid points in longitude due to regridding of the reconstruction. We define $Q$ as the longitudinal extent of the region influenced by CMEs. These CMEs are typically manifested as a brightening in the mass distribution map (\eg~No.~11 for COR1-A and No.~2 for COR1-B), but sometimes CMEs also cause the destruction of large coronal structure forming a long-lived coronal dimming (\eg~No.~5 and 8 for COR1-B). We estimate the change of coronal mass ($M_{\rm CME}$) caused by a CME by integrating the mass distribution over a region of size ${Q}\times{W}$ centered at the location ($L_{\rm c}$, $B_{\rm c}$) by first removing the mass profile derived from the pB data observed immediately prior to the CME (see the pre-CME time $t_{\rm preCME}$ in Table~\ref{tab:cme}). Here $Q$=6 degrees ({\it i.e.} covering 3 grid points in longitude) is assumed. The obtained values of $M_{\rm CME}$ are found in the range from $-3\times10^{13}$ g to $4\times10^{14}$ g (see Table~\ref{tab:cme}), where the positive and negative signs correspond to mass increase and decrease, respectively. If taking these CME-caused mass changes as errors for the total coronal mass ($M_{\rm total}\approx7\times10^{16}$ g) calculated for CR 2136 using Equation (\ref{equ:mtt}), we then derive a total error of only 0.20\% for COR1-A and 0.67\% for COR1-B. This result suggests that the uncertainty caused by CMEs in measurements of the total coronal mass from COR1 is negligible. The reason could lie in the fact that most of the CMEs (with their carried mass) originate from the low corona below 1.5 R$_\odot$ \citep[\eg][]{wan02, gib06}. Our suggestion is also supported by the recent study by \citet{lop17} who estimated for three CMEs both the CME mass and the low-corona evacuated mass and found them both to be of order $10^{15}$ g, with the latter explaining a large fraction of the former. However, sometimes when a CME blows out streamers (\eg~CME No.~8 for COR1-B due to a large filament eruption) or if the streamer itself erupts to become a CME (\eg~No.~5 and 10 for COR1-B), then the resultant mass loss of the corona could be large. \citet{kra11} analyzed such an event based on tomographic reconstructions of the 3D electron density in the corona before and after the CME using COR1 data and found a mass loss of $\approx1.0\times10^{15}$ g.

\begin{sidewaystable}
\caption{ CMEs observed in the COR1 pB images that are used to make the 3D density 
reconstruction of CR 2136 during the period 1--15 May 2013$^{a}$.}
\label{tab:cme}
\newcommand*{\head}[1]{\multicolumn{1}{c}{\scriptsize #1}}
\begin{tabularx}{\textheight}{lllrrrrcllrrrr}
\hline
  & \multicolumn{6}{c}{COR1-A} & & \multicolumn{6}{c}{COR1-B} \\
    \cline{2-7} \cline{9-14} \\
CME & $t_{\rm CME}$ & $t_{\rm preCME}$ & \head{$L_{\rm c}$} &  \head{$B_{\rm c}$} & \head{$W$} &  \head{$M_{\rm CME}$} &    & $t_{\rm CME}$ & $t_{\rm preCME}$ & \head{$L_{\rm c}$} &  \head{$B_{\rm c}$} & \head{$W$} & \head{$M_{\rm CME}$} \\
 No.& (UT) & (UT) & \head{(deg)} & \head{(deg)}  & \head{(deg)} & \head{(10$^{12}$ g)} & & (UT) & (UT) & \head{(deg)} & \head{(deg)}  & \head{(deg)} & \head{(10$^{12}$ g)}\\
\hline
1 & 0502 06:00 & 0502 05:00 & 213 &  45 &  70 &    6.3 & & 0502 06:00 & 0502 05:00 & 116 & 60  &  60 &    22.4 \\
2 & 0503 18:00 & 0503 17:35 & 14 &  35 &  63 &   52.7 & & 0503 18:00 & 0503 17:35 &  96 & 20  &  71 &    83.8 \\
3 & 0505 00:05 & 0504 23:10 & 357 & -28 &  40 &   23.0 & & \multicolumn{6}{c}{not seen} \\
4 & \multicolumn{6}{c}{not seen}          & & 0506 11:55 & 0506 10:05 &  60 & -37 &  48 &   $-$4.3 \\
5 & 0507 12:00 & 0507 09:20 & 324 & -75 &  32 & $-$6.5 & & 0507 11:55 & 0507 09:15 & 227 & -65 &  44 &  $-$3.9 \\
6 & \multicolumn{6}{c}{not seen}          & & 0508 00:05 & 0507 22:45 &  40 &  19 &  17 &      7.5 \\
7 & \multicolumn{6}{c}{not seen}          & & 0508 12:00 & 0508 10:50 & 214 &  -3 &  76 &    412.8 \\
8 & \multicolumn{6}{c}{not seen}          & & 0509 18:00 & 0509 17:10 & 197 & -39 &  42 &      6.4 \\
9 & 0510 17:55 & 0510 16:55 & 102 &  23 &  45 &   45.5 & & \multicolumn{6}{c}{not seen} \\
10 & \multicolumn{6}{c}{not seen}         & & 0511 06:00 & 0510 23:30 & 357 &  65 &  43 &  $-$33.0\\
11 & 0512 00:05 & 0511 22:25 & 85 &  35 &  50 &  115.9 & & 0512 00:05 & 0511 22:25 & 347 &  58 &  54 &     6.8 \\
12 & \multicolumn{6}{c}{not seen}         & & 0513 12:00 & 0513 06:00 & 148 & -54 &  41 &     48.8 \\
13 & 0513 12:00 & 0513 07:25 & 246 &  -3 &  25 &  37.1 & & 0513 12:00 & 0513 04:00 & 328 & -30 &  31 &    43.4 \\
14 & 0513 18:00 & 0513 17:10 & 62 &  13 &  65 &  53.3 & & 0513 18:00 & 0513 17:10 & 144 &  19 &  57 &    62.7 \\ 
15 & \multicolumn{6}{c}{not seen}         & & 0515 00:05 & 0514 22:35 & 128 & -20 &  60  &   111.2 \\
\hline
\end{tabularx}
\begin{tablenotes}
\small
\item $^a$ $t_{\rm CME}$ is the observing time of the pB images that capture a CME. $t_{\rm preCME}$ is the observing time of the pB images immediately previous to the CME. $L_{\rm c}$ and $B_{\rm c}$ are the Carrington longitude and latitude of the CME center position. $W$ is the latitudinal width of the CMEs, and $M_{\rm CME}$ is the CME-caused coronal mass change.
\end{tablenotes} 
\end{sidewaystable}

Since COR1-A and COR1-B generally observe the same coronal structure at different times (except when their separation is close to $0^\circ$ or $180^\circ$), we may estimate errors of the coronal mass due to temporal evolution (including destruction of streamers) based on the difference of mass distributions between COR1-A and COR1-B. As an example, consider the 3D density reconstruction for CR 2136. Figure~\ref{fig:cme} shows that COR1-B observed a dimming region (at Carrington longitude from 160$^\circ$ to 195$^\circ$ and latitude from $-60^\circ$ to $-20^\circ$) following CME No.~8 (see panel (b)), while COR1-A observed the pre-erupted coronal structure about 6 days before the CME (because the separation between COR1-A and -B was 83$^\circ$). Thus the mass loss in the dimming region can be calculated from the difference of mass distributions between COR1-A and COR1-B. This example also implies that when we use the mean mass distribution between COR1-A and -B, the errors due to temporal evolution can be reduced by $\approx$50\%. To be more general, we define ``significant changes" in the mass distribution due to temporal evolution as the unsigned mass differences between COR1-A and -B above 2$\sigma$. The $\sigma$ here is the standard error for the average of $\Delta{M}(i,j)=M_A(i,j)-M_B(i,j)$, where $M_A(i,j)$ and $M_B(i,j)$ are the mass distributions for COR1-A and -B, respectively, obtained using Equation~(\ref{equ:mss}). Figure~\ref{fig:dif}a shows the regions ($S$) with significant mass change (enclosed with the contour) for CR 2136, which cover the dimming region mentioned above. By taking the total unsigned difference within region $S$ as an estimate of uncertainty in the total coronal mass caused by temporal evolution, we derive the relative error $r_{\rm error}^M\approx$5\% for this reconstruction using the expression
\begin{equation}
r_{\rm error}^M=\frac{\sum_{i,j\in{S}}|M_A(i,j)-M_B(i,j)|}{M_A+M_B}, ~~~~(S=S(|\Delta{M}(i,j)|>2\sigma)),  \label{equ:erm}
\end{equation}
where $M_A$ and $M_B$ are the total coronal mass for COR1-A and -B calculated using Equation~(\ref{equ:mtt}). With this method we estimate the errors for the 3D density reconstructions of CRs 2054--2153 (see the red line with pluses in Figure~\ref{fig:err}b), and find that $r_{\rm error}^M$=1--10\%  with mean values of 3.4\% and 5.1\% during the minimum phase and the maximum phase, respectively. 

In addition, we compare the radial dependence of globally-averaged electron densities between COR1-A and -B for CR 2136 (see Figure~\ref{fig:dif}b), and find that they are consistent over the lower (1.5--2.6 R$_\odot$) region of large density while their difference becomes distinct at the higher region close to the outer boundary of the FOV where the signals are weak. This feature suggests that the instrumental background noise may be an important source for uncertainty in the low density region. To estimate the error for the globally-averaged radial density distribution we calculate the root mean square of the normalized density difference between COR1-A and -B using
\begin{equation}
r_{\rm error}^N=\sqrt{\frac{1}{n}\sum_{k=1}^{n}\left(\frac{N_A(r_k)-N_B(r_k)}{N_A(r_k)+N_B(r_k)}\right)^2}, \label{equ:ern}
\end{equation}
where $N_A(r_k)$ and $N_B(r_k)$ are the globally-averaged radial density profiles for COR1-A and -B, respectively, and $n=23$ is the total number of radial grid points. We obtain $r_{\rm error}^N\approx$17\% for CR 2136. For the 3D reconstructions of CRs 2054--2153 we find that $r_{\rm error}^N$=3--26\% with mean values of 14\% and 16\% during the minimum and maximum phases, respectively (see the green line with crosses in Figure~\ref{fig:err}b). We also find that except for the period from 2008 to 2009 and at several peaks of $r_{\rm error}^M$ (\eg~in early 2010 and early 2012), the errors $r_{\rm error}^N$ and $r_{\rm error}^M$ vary with time roughly in the same trend. Noticeably the $r_{\rm error}^N$ drops about 50\% after January 2009. This is most likely due to the serious dust deposition event on 30 January 2009 that led to the COR1-B background increasing from a previously much lower level to that comparable to COR1-A \citep{tho10}. 

The subtraction of the instrumental stray light background is an important step in the COR1 calibration. The routine {\scshape secchi\_prep} in SSW processes COR1 images with a choice of two types of background images: the regular monthly minimum backgrounds (by default) or the combined backgrounds from both monthly minimum and calibration roll images (with the keyword {\scshape /calroll}) (see \inlinecite{tho10} for details). The purpose of calibration rolls is to improve the background images by rejecting the residual K-coronal light from persistent streamers (mostly in the equatorial regions). Here, we apply the combined backgrounds to all pB images used in the density reconstructions. In the following, we analyze the uncertainty caused by the background subtraction based on differences between the total coronal masses from COR1-A and -B. Figure~\ref{fig:err}b shows that the relative mass differences ($r_{\rm dif}=|M_A-M_B|/(M_A+M_B)$; see the black line with pluses) have several peaks above 10\%. The biggest peak between early 2007 and 2008 results from the absence of calibration rolls during this period while the other peaks may be involved with the events that affected the COR1 background subtraction. Such events include the spacecraft repoint, the image binning format change, the exposure time change, and the deposition of dust particles on the objective lens (see \opencite{tho10}). Particularly, the dust landing events cause a sudden jump in the scattered light background, which is followed by some slight decrease at the beginning. Because the background data are treated separately before and after each event (by calling the routine {\scshape scc\_getbkgimg}), the background subtraction close to the event does not work well, especially when the combined background with calibration rolls is applied. By comparing with the results re-calculated from the 3D density reconstructions made with the pB images processed with the regular monthly minimum backgrounds (see Figure~\ref{fig:reg} in the Appendix), we confirm that several big peaks in $r_{\rm dif}$ are due to use of calibration roll backgrounds. However, applying no calibration roll backgrounds leads to an underestimation of the total coronal mass by $20\pm8$\% on average (see Figure~\ref{fig:rat} in the Appendix). In addition, the calibration roll background may not work well during solar maximum because of the presence of polar streamers (or absence of CHs). Finally, we notice a systematic drop of $\approx$10\% in the coronal mass evolution after 19 April 2009 due to the change of image binning format (see an arrow marked in Figure~\ref{fig:err}a). This could be attributed to an alternate way the data were compressed for telemetering. 

Finally, we emphasize that the local spherical symmetry assumption which the SSPA technique is based on, while valid for specific observations (\eg~streamer edge-on view), it is not valid in general (\eg~streamer face-on view). The effect of this situation was demonstrated in Figures 7 and 8 of \citet{wan14}, where it was shown that SSPA is able to recover the radial electron density profile of a tomographic model of streamers when the favorable viewing conditions are met so that the symmetry in longitude is roughly valid. It was also shown that even in such cases their matching degree decreases with height as the streamer region takes a progressively smaller part of the LOS. The quantitative comparisons of SSPA reconstructions with the tomographic density model and the MHD density model in this paper show that the uncertainty of SSPA is within a factor of about 2 for the most regions where their density ratios are between 1/2 and 2 (see right panels of Figure~\ref{fig:ntom} and Figure~\ref{fig:nmod}). These comparisons also show that the SSPA and model densities appear to agree better at higher height. Despite some differences in fine coronal structures between the SSPA and model reconstructions, their globally-averaged radial density profiles are consistent (see Figures~\ref{fig:ntmr} and~\ref{fig:nmdr}). This suggests that the spherical symmetry approximation may affect the density reconstruction like ``smoothing" which only causes the redistribution of coronal mass (mainly along longitude) but does not change the total mass. Figure 9 of \citet{wan14} showed such an instance, where the difference of the total mass integrated over 1.6$-$3.8 R$_{\odot}$ between the SSPA inversion and the given density model was about 7\%. This smoothing effect can also be verified based on the 2D toy models given in the Appendix of \citet{wan14}. 

Here we calculate the total mass in the 1.5$-$3.7 R$_{\odot}$ region using the SSPA and tomographic reconstructions for CR 2066, and obtain $M_{\rm total}^{\rm SSPA}=5.5\times10^{16}$ g for COR1-B and $M_{\rm total}^{\rm tomo}=3.9\times10^{16}$ g for the tomographic model, where we compare $M_{\rm total}^{\rm SSPA}$ for COR1-B with $M_{\rm total}^{\rm tomo}$ as the tomographic reconstruction was made of the COR1-B data. We find that the difference between the total mass is $\approx$39\%, which reduces to, however, only $\approx$10\% when integrated over the 1.7$-$3.7 R${\odot}$ region. The larger difference in the former case is mainly due to the fault of the tomographic model near the occulter (see Figure~\ref{fig:ntmr}). We also compare the total mass calculated in 1.5$-$3.7 R$_{\odot}$ between the SSPA reconstruction and the MHD density model for CR 2097/2098, and find that $M_{\rm total}^{\rm SSPA}=5.3\times10^{16}$ g and $M_{\rm total}^{\rm MHD}=5.1\times10^{16}$ g, which differ by about 4\%. There is another caveat that can be attributed to the above two examples of SSPA reconstructions, {\it i.e.} they are made for CRs during solar minimum or the early rising phase, whose density structures during that period are relatively simple and stable. It is known that  coronal structures are more complicated and dynamic during solar maximum, and therefore, a similar analysis of uncertainty for solar maximum rotations (\eg~CR 2136 shown in Figure~\ref{fig:cme}) is required in a future study. 

\section{Coronal Activity}
\label{sctana}
\subsection{Long-Term Variations}
\label{sctvar}
\begin{figure}
\centerline{\includegraphics[width=0.9\textwidth,clip=]{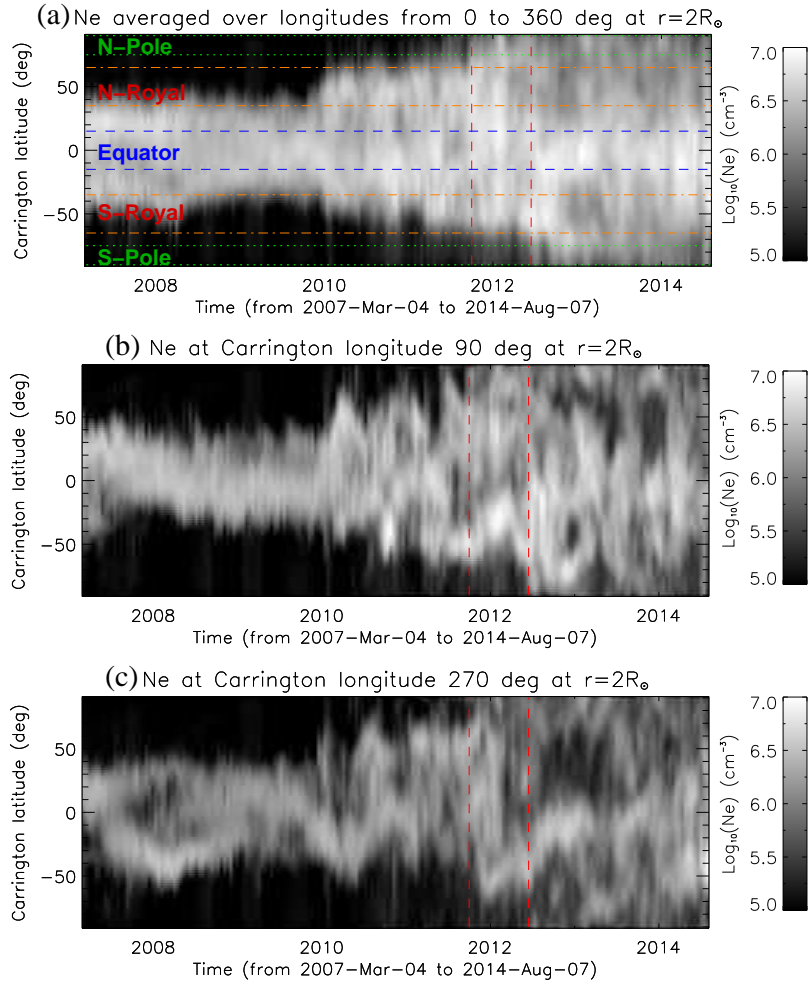}}
\caption{Time evolution of latitudinal distributions of the mean electron density for COR1-A and -B at a heliocentric distance of 2 R$_\odot$, (a) averaged over all longitudes, (b) along the cut at longitude 90$^{\circ}$, and (c) along the cut at longitude 270 $^{\circ}$. Two vertical dashed lines mark the time when streamers reach the North Pole (in October 2011) and the South Pole (in June 2012), respectively. Horizontal lines in (a) delineate polar zones (dotted lines), royal zones (dot-dashed lines), and the equatorial zone (dashed lines). }
\label{fig:nlat}
\end{figure}

We use the 3D electron density reconstructions for CRs 2054--2153 to study the temporal evolution of the global corona. Figure~\ref{fig:nlat} shows three time-latitude maps of the electron density, made by stacking in time the longitude-averaged densities (panel (a)), the cut at 90$^\circ$ longitude (panel (b)), and the cut at 270$^\circ$ longitude. The streamer belt is concentrated in the equatorial region during the minimum period of solar activity (from 2007 to 2009), and then expands toward the polar regions as the level of activity increases. Finally, it reaches the polar regions around 2012 and persists globally during the maximum period of solar activity (from 2012 to 2014). A careful examination finds that streamers reached the North Pole in October 2011; about 8 months earlier than they reached the South Pole. The behavior of the streamer belt is closely related to temporal evolution of the magnetic neutral sheet or the HCS \citep{sch73, guh96, sae07, hu08}. Its shape gets progressively deformed from a rather flatter plane (concentrated in the equatorial band) around solar minimum to a highly warped surface (reaching high-latitude regions) at solar maximum. 

\begin{figure}
\centerline{\includegraphics[width=1.\textwidth,clip=]{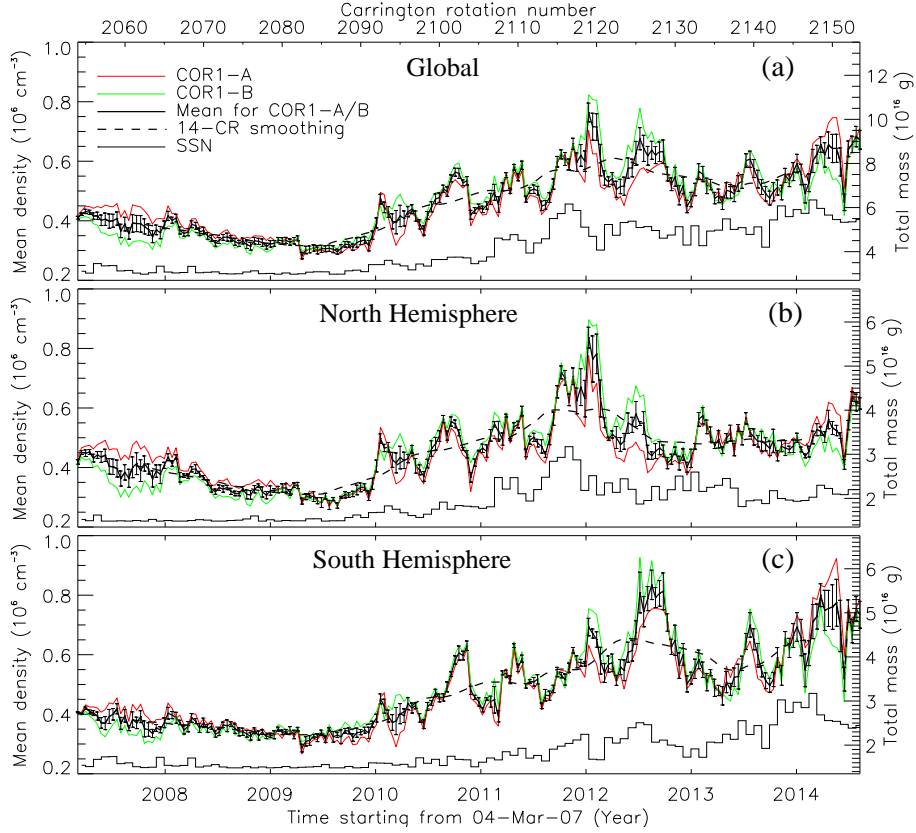}}
\caption{Temporal variations of the electron density in the corona, averaged from 1.5 to 3.7 R$_\odot$ globally (a), in the northern hemisphere (b), and in the southern hemisphere (c). The red line represents the densities derived from STEREO/COR1-A data, and the green line represents the densities from COR1-B data. The thick black line with the $+$ symbols represents the mean densities for COR1-A and -B, and the thick dashed line corresponds to the 14-CR ($\approx$13-month) running average. The thin black line in histogram mode represents the Wolf sunspot numbers (SSN) integrated over each CR (shown in arbitrary unit). The right $y$-axis indicates the corresponding total coronal mass integrated from 1.5 to 3.7 R$_\odot$ globally (a), and in hemispheres (b) and (c).} 
\label{fig:nvar}
\end{figure}

We calculate the total mass of the global corona ($M_{\rm total}$) from the 3D density reconstructions using Equation~(\ref{equ:mtt}). By applying $M_{\rm total}=\mu{m_p}N_{\rm total}$, we then derive the globally-averaged electron (number) density,
\begin{equation}
N_{\rm mean}=\frac{N_{\rm total}}{V_{\rm total}}=\frac{M_{\rm total}}{\mu{m_p}V_{\rm total}}=C{M_{\rm total}},
\end{equation}
where $N_{\rm total}$ is the total number of electrons in the analyzed spherical region between $r_1$=1.5 and $r_2$=3.7, which has the total volume $V_{\rm total}=(4/3)\pi R_\odot^3(r_2^3-r_1^3)$, and where the constant $C=7.5\times10^{-12}$ cm$^{-3}$g$^{-1}$. As the total mass and the global mean electron density of the corona only differ by a constant factor, their evolution is shown using the same curve (see Figure~\ref{fig:nvar}a). Likewise, the calculated total mass and mean electron density in the northern and southern hemispheres are shown in Figures~\ref{fig:nvar}b and~\ref{fig:nvar}c, respectively. The reason for considering the two hemispheres separately lies in the hemispheric asymmetry of magnetic activity such as the inequality of sunspot numbers \citep[\eg][]{mci13,bis14,ben14}. In comparison, the total Wolf sunspot number (SSN) integrated over each CR is overplotted. The daily total and hemispheric SSN data are publicly available on the WDC-SILSO archive\footnote{http://sidc.oma.be/silso/datafiles}. 

Figure~\ref{fig:nvar} shows that the long-term trend and overlying short-term oscillations of coronal mass variations roughly follow the behavior of the sunspot number implying dependence on the magnetic activity evolution on the solar surface. The two hemispheres show that the oscillations are clearly out of phase. The measurement shows that the northern oscillation is leading the southern oscillation during the rising phase by $\approx$7 months (8 CRs), based on the time lag between their maxima (reached at about 2012/01 in the northern hemisphere and at 2012/08 in the southern one). Note that this time lag is close to the time difference for the streamer belt reaching the northern and southern poles. The variability of the K-corona is often characterized by the so-called modulation factors (MFs) that are defined as the ratios between the maximum and minimum of the integrated radiance or pB \citep[\eg][]{fis84, bar15}. To determine MFs for the temporal variation of total coronal mass (or mean electron density), we first calculate the 14-CR ($\approx$13-month) running averages that represent the long-term variations (see the dashed lines in Figure~\ref{fig:nvar}), and then measure their minimum and maximum values. The 13-month running average is a standard smoothing method, which is widely used (see \opencite{hat15}). We obtained MF= 1.9, 1.9, and 2.0 for the global, the northern and southern hemispheric corona, respectively. The modulation factors indicate that the variation amplitude in the southern hemisphere is slightly larger than in the northern hemisphere.

\begin{figure}
\centerline{\includegraphics[width=1.\textwidth,clip=]{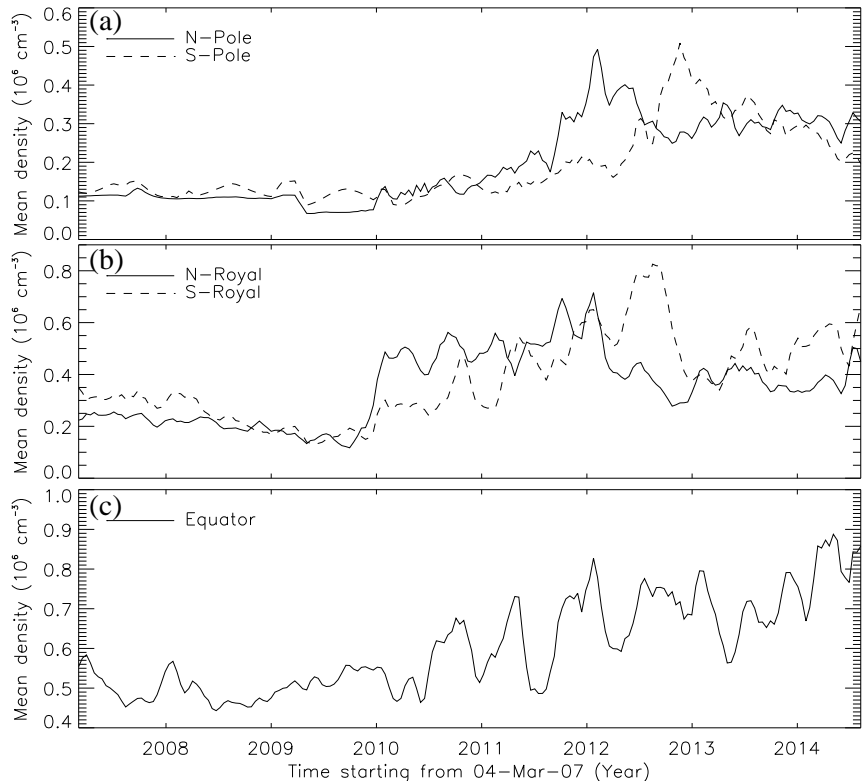}}
\caption{Temporal variations of the electron density averaged from 1.5 to 3.7 R$_\odot$ in different regions. (a) North and south polar zones. (b) North and south royal zones. (c) Equatorial zone. All curves in (a)--(c) are calculated as averages for COR1-A and -B, and then smoothed with a 2-pixel (1-CR) running average.}
\label{fig:npre}
\end{figure}

Some studies revealed that the hemispheric asymmetry was latitudinally dependent \citep[\eg][]{bis14, bar15}. To analyze this behavior we define three latitude regions similar to those in \citet{bar15}: the North or South Pole in latitude 75$^\circ$--90$^\circ$, the north or south high-latitude region from 35$^\circ$--65$^\circ$ (also called royal zones, see the definition in Figure 1 of \inlinecite{bar15}), and the equator within latitudes $\pm15^\circ$ (see Figure~\ref{fig:nlat}a). We compare the temporal variations of the coronal average density in these regions (see Figure~\ref{fig:npre}). We find that two royal zones vary with a phase difference that is comparable to the two hemispheres, showing that the northern zone leads the southern zone by $\approx$7 months (a time shift between their maximum peaks). A distinct phase difference is also observed between the two polar zones. We measure a time lag of $\approx$9 months between their maxima (reached at 2012/2 in the North Pole and at 2012/11 at the South Pole). The short-term fluctuations are obvious at the equator during the rising and maximum phases of Cycle 24, while relatively weaker in the poles during this period. From the 14-CR running averages we determine the long-term variations of the average density (or total mass) in different zones in terms of MF. The measured values are listed in Table~\ref{tab:mf}. The modulation factors indicate that the strongest variation is in the polar region, while the weakest is at the equator. In addition, the MF for the southern royal zone is larger than that for the northern royal zone and is  consistent with the case for the two hemispheres.   

\begin{figure}
\centerline{\includegraphics[width=1.\textwidth,clip=]{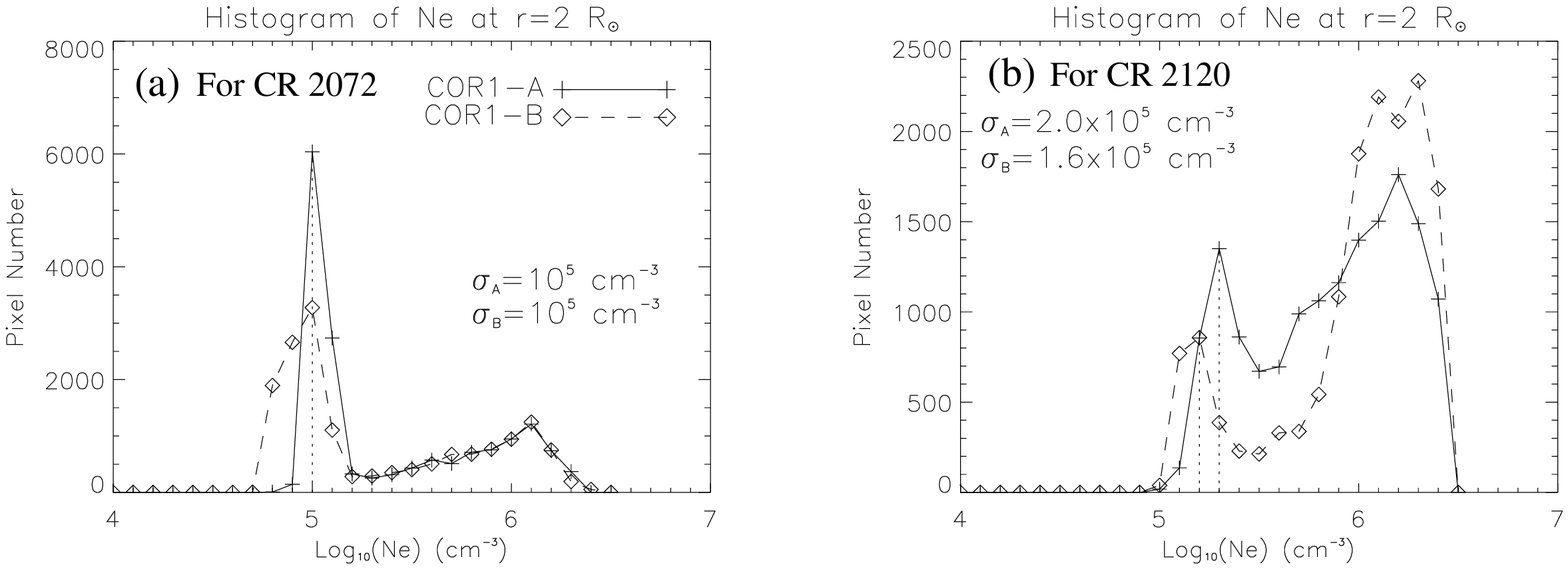}}
\caption{Histograms of the coronal electron densities at a heliocentric distance of 2 R$_\odot$ for CR 2072 (a) and CR 2120 (b). The solid lines with crosses represent the densities for STEREO/COR1-A, and the dashed lines with diamonds for COR1-B. The background noise ($\sigma$) of the 2D density distributions (as shown in Figure~\ref{fig:nemap}) is estimated as the density value at the first frequency peak (marked with the vertical dotted lines). The determined noise levels for COR1-A and -B (with $\sigma_A$ and $\sigma_B$) are also marked on the plots. }
\label{fig:nsgm}
\end{figure}

\begin{figure}
\centerline{\includegraphics[width=1.\textwidth,clip=]{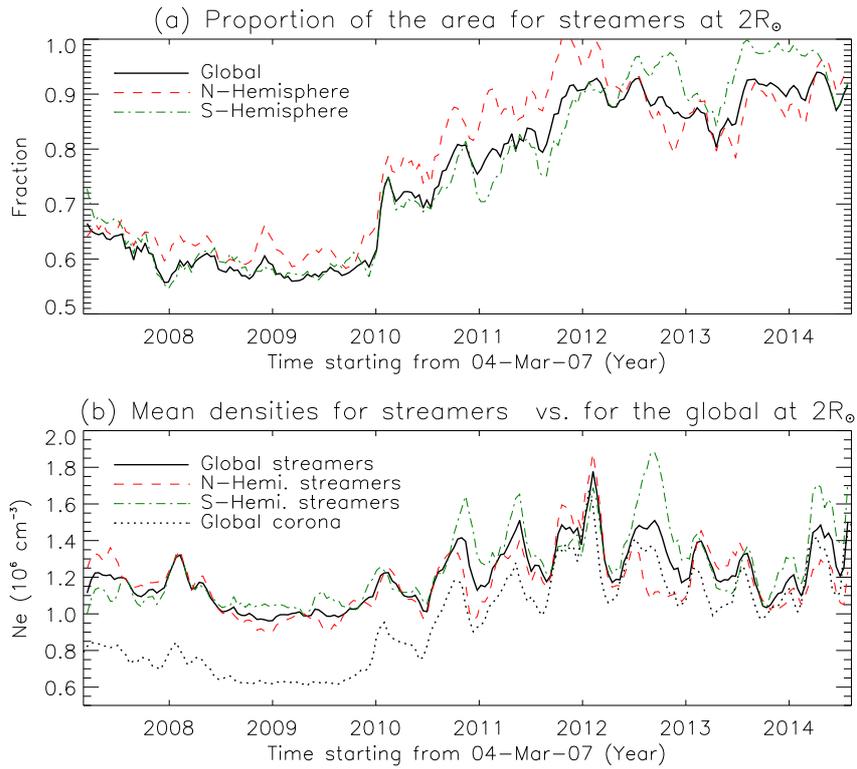}}
\caption{(a) Proportions of the area of the streamer region (with $N_{\rm e}>3\sigma$) in the global corona (thick solid line), the northern hemisphere (dashed line) and the southern hemisphere (dot-dashed line) at $2 R_\odot$. (b) Temporal variations of the average density for the streamer region in the global corona and the two hemispheres. In comparison, the dotted line indicates the globally averaged density (or the total mass) at $r=2 R_\odot$. All curves in (a) and (b) are calculated as the averages for COR1-A and -B, and then smoothed with a 2-pixel (1-CR) running average.}
\label{fig:narv}
\end{figure}

\subsection{Variation of Streamers}
\label{sctstr}
\begin{figure}
\centerline{\includegraphics[width=1.0\textwidth,clip=]{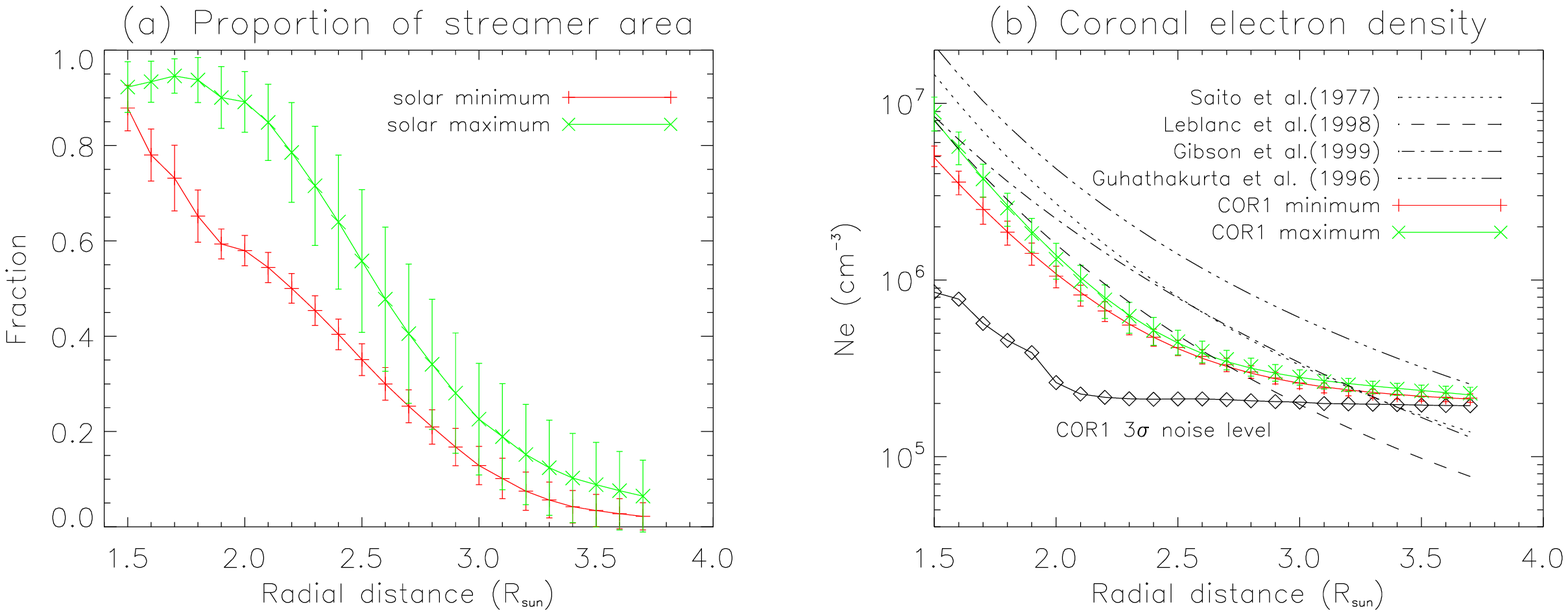}}
\caption{(a) Proportion of the total area of streamer regions (with $N_{\rm e}>3\sigma$) in the spherical cross section of the 3D density as a function of radial distance. The red line with pluses represents the radial distribution averaged over CRs 2064--2089 for both COR1-A and -B during solar minimum, while the green line with crosses represents the average over CRs 2114--2152 for COR1-A and -B during solar maximum. Error bars are the standard deviation for the corresponding average. (b) Radial distributions of the streamer electron density averaged over the areas of $N_{\rm e}>3\sigma$ for COR1-A and -B. The red solid line is a least-square polynomial fit to the radial average densities (denoted with pluses) at solar minimum, while the green solid line is a fit to the radial average densities (denoted with crosses) at solar maximum. The black solid line with diamonds represents the COR1 background noise level (3$\sigma$) as a function of radial distance. For comparison, the density models of \citet{sai77}, \citet{leb98}, \citet{gib99}, and \citet{guh96} are overplotted as the dotted line, dashed line, dot-dashed line, and the dot-dot-dashed line, respectively. }
\label{fig:nnrv}
\end{figure}

  We analyze temporal variations of the total mass in the global corona and in the two hemispheres in the last section. As most of the coronal plasma (or electron content) concentrates in streamers, the analysis is indeed equivalent for the whole streamers (including the streamer belt, pseudo and polar streamers). In this section, we analyze temporal variations of the total area and the average electron density of the whole streamers using the SSPA 3D density reconstructions. We define the streamer region as the location where the densities are above the 3$\sigma$ noise level. The background noise $\sigma$ is estimated as the density value at the first peak in the histogram of the densities (in logarithm) created from the spherical cross-sectional density map. Figure~\ref{fig:nsgm} illustrates the determination of $\sigma$ from the density maps at 2 R$_\odot$ for CR 2072 and CR 2120. The corresponding streamer regions obtained from the 3$\sigma$ criteria are shown in Figure~\ref{fig:nemap} (enclosed with the contours). As the background noise only weakly depends on time, we simply fix noise levels at different heights to measure the variation of the streamer regions. The fixed noise levels are taken as the averages over 26 CRs at solar minimum (from CR 2064 to CR 2089). The black solid line (with the diamond symbols) in Figure~\ref{fig:nnrv}b indicates the radial dependence of the 3$\sigma$ noise level averaged for COR1-A and -B.

The measured total areas of streamer regions in the global corona and in the two hemispheres at 2 R$_\odot$ for CRs 2054--2153 are shown in Figure~\ref{fig:narv}a. The global streamer area increases from below 60\% at Cycle 23/24 minimum to above 90\% of the whole corona at Cycle 24 maximum. The streamer area in the northern hemisphere is larger than that in the southern hemisphere and is leading in phase during the rising period of solar activity, while it dominates in the southern hemisphere during the maximum phase. From the 14-CR running averages of the time profiles, we measure the modulation factors for the total streamer area to be MF= 1.6--1.7 (see Table~\ref{tab:mf}). The values for the two hemispheres are very close.  Figure~\ref{fig:narv}b shows temporal variations of the average density of streamer regions in the global corona and in the two hemispheres at 2 R$_\odot$. We measure the modulation factors from their 14-CR running averages and compare them with those of the globally- and hemispherically-averaged coronal densities at 2 R$_\odot$ (equivalent to the total mass; see Section~\ref{sctvar}). These measurements are listed in Table~\ref{tab:mf}. The modulation factors of the latter (MF=2.1--2.2) are clearly larger than the former (MF=1.4--1.5). This can be apparently explained by the fact that the increase of the total coronal mass results from increases in both the total area in 2D (or volume in 3D) and the average density of streamer regions from the minimum to maximum phase. In addition, the modulation factors of the streamer average density in the two hemispheres are very close. We also find that the oscillations of the streamer average density in the two hemispheres are correlated, and the peaks in the southern hemisphere appear to be larger in amplitude. Finally, it is noted that the oscillation of the streamer average density (thick solid line) is well correlated with that of the globally-averaged density (thick dotted line), and that the amplitude of the streamer density oscillation is much larger than that of the streamer area oscillation. These facts may suggest that the oscillations in the global coronal mass (see Figure~\ref{fig:nvar}a) are mainly due to the oscillations in the streamer density. 
 
Figures~\ref{fig:nnrv}a and~\ref{fig:nnrv}b show the radial dependence of the total area and the average density of streamer regions at solar minimum and solar maximum, respectively. The minimum-phase distributions are calculated by averaging over CRs 2064--2089 and the maximum-phase ones are calculated by averaging over CRs 2114--2152. The proportion of the total streamer area to the whole spherical area (at the same height) decreases with the radial distance at both solar minimum and maximum, but it appears to decrease faster at minimum than at maximum over the radial distance ranging from 1.5--2.0 R$_\odot$. The ratio of the total streamer area at solar maximum to that at minimum at the radial distance ranging from 1.5--3.7 R$_\odot$ is within 1.0--2.9 with a mean of 1.8. We fit the radial profiles of the average streamer density to a 4th-degree polynomial of the form    
\begin{equation}
N(r)=\frac{a_1}{r}+\frac{a_2}{r^2}+\frac{a_3}{r^3}+\frac{a_4}{r^4}, ~~~ (1.5 {\rm R}_\odot\leq{r}\leq 3.7 {\rm R}_\odot).
\end{equation}
We obtain $a_1=(0.4\pm2.0)\times10^6$, $a_2=(1.0\pm1.6)\times10^7$, $a_3=-(5.5\pm4.2)\times10^7$, and $a_4=(8.2\pm3.5)\times10^7$ for the solar minimum profile ($N(r)_{\rm min}$), while $a_1=-(4.4\pm3.7)\times10^6$, $a_2=(5.4\pm3.0)\times10^7$, $a_3=-(1.9\pm0.8)\times10^8$, and $a_4=(2.1\pm0.7)\times10^8$ for the solar maximum profile ($N(r)_{\rm max}$). The ratio $N(r)_{\rm max}/N(r)_{\rm min}$ decreases from 1.8 to 1.1 with increasing radial distance  from 1.5 to 2.6 R$_\odot$ with a mean of 1.3 in this range. For comparison, Figure~\ref{fig:nnrv}b also includes the plots for some previous coronal density models obtained in (or near) solar minimum. The dotted curve corresponds to the \citet{sai77} density model for the equatorial background (when no streamers or holes were visible), and the dot-dashed curve is the \citet{gib99} density model for streamers. Both density models were obtained from pB observations using the VdH method \citep{van50}. The dashed curve is a coronal electron density model derived from radio observations of type III bursts \citep{leb98}. We find that $N(r)_{\rm max}$ is consistent with the \citet{leb98} density model in the 1.5--3.0 R$_\odot$ range. The \citet{sai77} density distribution in the 1.5--3.0 R$_\odot$ range is on average larger than $N(r)_{\rm max}$ by a factor of $\approx$1.8 and larger than $N(r)_{\rm min}$ by a factor of $\approx$2.2. The \citet{gib99} density model is in between the \citet{leb98} and \citet{sai77} models. It is noticed that the average streamer density $N(r)$ for COR1 distinctly deviates from the \citet{leb98} density model beyond 3 R$_\odot$. This is caused by our definition of ``streamer regions" satisfying $N_{\rm e}>3\sigma$ where $3\sigma\approx2\times10^5$ cm$^{-3}$ in the range 3.0--3.7 R$_\odot$. So $N(r)$ can only approach to the 3$\sigma$ level when decreasing with $r$ but will never fall below this value. Thus it is reasonable to restrict the application range of the obtained density function $N(r)$ to the region 1.5--3.0 R$_\odot$. In addition, the \citet{guh96} density model (the dot-dot-dashed line) is also overplotted in Figure~\ref{fig:nnrv}b, which is derived based on the same calibrated data set (from {\it Skylab}) as analyzed by \citet{sai77} but for different coronal structures. The \citet{guh96} density model was obtained at the current sheet (taken as the center or the brightest location of the streamer belt), so the density values may be regarded as an upper limit for streamers. We find that the average density at the current sheet obtained by \citet{guh96} is a factor of 3.6 higher than that of $N(r)_{\rm min}$ over the range 1.5--3.0 R$_\odot$ for the streamer region as defined here.  

\subsection{Short-Term Variations}
\label{sctsvr}
\begin{figure}
\centerline{\includegraphics[width=1.\textwidth,clip=]{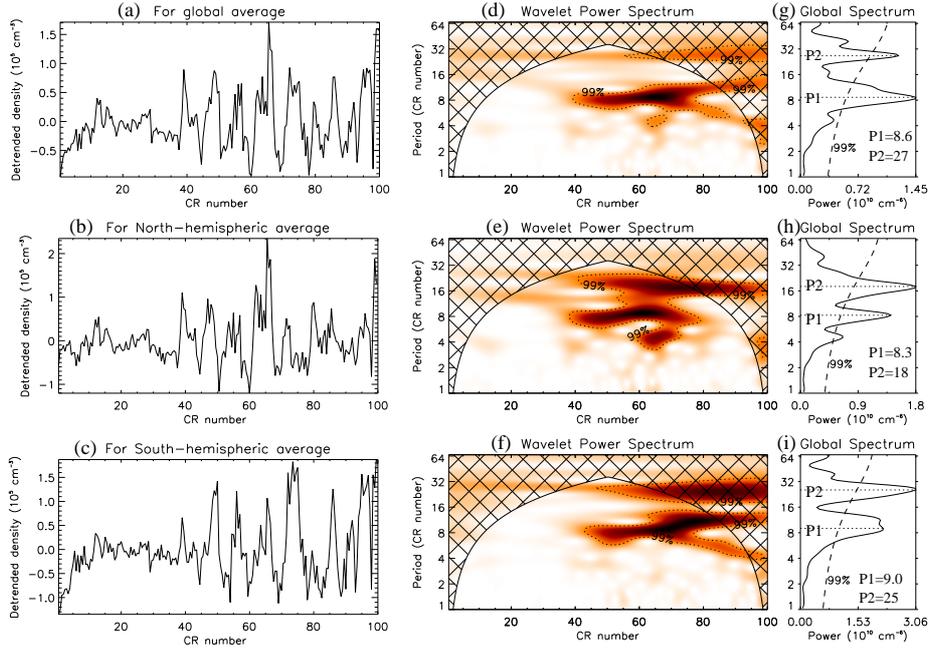}}
\caption{(a)--(c) Detrended electron density variations over the global, north-hemispheric, and south-hemispheric corona, respectively. (d)--(f) Corresponding wavelet power spectra. The dark color represents high power in the power spectra and the dotted contour encloses regions of greater than 99\% confidence. The black grid indicates the region where estimates of oscillation period become unreliable. (g)--(i) Corresponding global wavelet power spectra. Peaks (in solid line) above the 99\% confidence level curve (in dashed line) are statistically significant. }
\label{fig:nwav}
\end{figure}

As mentioned above, the average electron density (or total mass) of the global and hemispheric corona shows clear quasi-periodic oscillations during the rising and maximum phases of Cycle 24. We now analyze these oscillations using the wavelet method \citep{tor98}. This method allows us to identify the periodic components in a time series and their variation with time. For the convolution of the time series the Morlet wavelet was chosen. The global wavelet spectrum (GWS) is the average of the wavelet power over time at each oscillation period. Statistically significant oscillation periods are defined here as exceeding the 99\% confidence level against the white noise background. In practice, we first subtract the slowly varying long-term trend from the time series. The trend is constructed using Fourier low-pass filtering with a cutoff period of $P_{\rm c}\gtrsim$20 CRs. The panels (a)--(c) of Figures~\ref{fig:nwav} show temporal variations of the detrended electron density averaged (for COR1-A and -B) over the region 1.5--3.7 R$_\odot$ of the global, north-hemispheric, and south-hemispheric corona, respectively. The panels (d)--(i) show their wavelet analyses. Although two main peaks in the GWS are statistically significant ($>$99\% confidence level) (see panels (g)--(i)), the long-period peak ($P_2$=18--27 CRs) has power mostly in the ``cone of influence" (see panels (d)--(f)), so the estimate of its oscillation period is not reliable due to the edge effects. Thus we determine the period of the short-term oscillations of the corona from the short-period peak ($P_1$=8--9 CRs, {\it i.e.} 7--8 months) in the GWS. Taking the uncertainty of the measurement as the FWHM of the GWS peak, we obtain $P_1=9\pm3$ CRs for the density variation of the global corona.

\section{Discussion and Conclusions}
\label{sctdc}
In this study, we reconstruct the 3D electron density models of the corona for CRs 2054--2153 using the SSPA method from STEREO/COR1 pB observations during 2007/3--2014/8. These 3D density reconstructions are validated by comparison with examples of similar models created by other methods such as tomography and MHD simulation as well as by 2D density distributions inverted using the VdH technique from LASCO/C2 pB images. Some previous studies confirmed that the Spherically Symmetric Inversion (SSI) method was applicable to the solar minimum streamer (belt), which gave the coronal densities consistent with those by other techniques such as spectroscopy and tomography within a factor of two \citep{gib99, lee08, wan14}. Here we also examine a solar maximum case. By comparing the electron density distributions of CR 2120 (in February 2012) with those inverted from the LASCO/C2 observations, we find that their uncertainties are of the same order (within a factor of two) as in the solar minimum case. This suggests that our SSPA 3D coronal density models reconstructed for 100 CRs (with a cadence of about two weeks) may be used for the interpretation of radio bursts (such as type II and moving type IV produced by CMEs) as observed from the Earth direction \citep[\eg][]{cho07, ram13, she13, sas14, har16, lee16}, in particular, when LASCO/C2 pB data are not available. We estimate the total mass (or electron content) contained in the coronal region observed with COR1 and its evolution with solar cycle. These measurements are important for use of testing different heating models for magnetic structures in the solar atmosphere \citep{lio09}. The error analysis suggests that the effect of CMEs is trivial while the temporal evolution, instrumental background subtraction, and the spherically symmetric approximation are the major sources of uncertainty in 3D reconstructions of the global corona and estimation of the global coronal mass by the SSPA technique.

We study the long-term variations of the global and hemispheric corona from solar minimum to the maximum of Cycle 24. A clear hemispheric asymmetry in the evolution of streamers and total coronal mass is found. During the rising phase the streamer (belt) expands from the equator towards high latitudes. The streamers reach the North Pole earlier than they reach the South Pole by $\approx$8 months. The variations of the coronal mass in the two hemispheres show a similar phase shift ($\approx$7 months) with the northern hemisphere leading. The further analysis of the latitudinal dependence of the north-south asymmetry shows that the phase difference between the two poles ($\approx$9 months) is similar to that between the two royal zones ($\approx$7 months). In contrast, the measurements for these regions based on variations of the K-coronal radiance using LASCO/C2 were divergent \citep{bar15}: a time lag of 1 month was found between the two hemispheres, 8 months between the two royal zones, and 17 months between the two poles. 
The discrepancy between the past results and that in this paper may be due to: i) different background subtraction techniques for LASCO/C2 and STEREO/COR1; ii) different definitions for the Cycle 24 rising phase; iii) different FOVs between C2 and COR1. First, the LASCO/C2 data calibration requires a sophisticated procedure in separating the K corona from the F corona and straylight (see \inlinecite{lle04,lle12} for details), where the morphology of C2 straylight is invariant during long periods of time, whereas this is not the case for STEREO/COR1 (see Table~\ref{tab:evt} and \inlinecite{tho10}). Our study is based on the COR1 pB data with {\scshape /calroll} background subtraction (see Section~\ref{scterr}). \citet{fra12} showed that, when used with the {\scshape /calroll} option, COR1 pB data matched well with measurements from LASCO/C2 within streamers, but the COR1 data were very low in other regions such as CHs. This is because the {\scshape /calroll} background subtraction method basically takes the CH data as the background level, and thus underestimates the pB. This underestimation affects the entire corona, but is more pronounced in the polar CHs because of their low brightness. The underestimation has a radial dependence, but is insensitive to position angle. Therefore, the underestimation applies equally to the northern and southern hemispheres, and thus should not affect any intercomparison between these two regions. However, the underestimation may become worse during solar maximum as a result of lack of CHs. This could explain the reason for the slight decrease in the measured total coronal mass during this period (see Figure~\ref{fig:nvar}). The second reason for the difference between the COR1 and LASCO/C2 results could be that we measure the north-south phase shifts based on the time difference of their largest peaks, while \citet{bar15} used a different technique. For example, we find that the South Pole reaches the maximum at 2012/11, whereas \citet{bar15} determined its maximum at about 2013/05 which corresponds to the second largest peak for the South Pole (see panel D of their Figure 2). In addition, \citet{bar15} did not mention how the hemispheric phase shift was obtained. The third reason could lie in the fact that the total coronal masses for COR1 and C2 are measured by integrating over different radial ranges due to different FOVs, which may lead to different variations. 

\begin{table}
\caption{ Modulation factors for the total mass, total area and average electron density of streamers as defined in the text and calculated from the 14-CR (13-month) running average. }
\label{tab:mf}
\begin{tabular}{lclc}
\hline
Regions\tabnote{For 3D regions from 1.5 to 3.7 R$_\odot$.}   &  MF  &  Regions  &  MF  \\
\hline
Global coronal mass in 3D       &    1.9    & Global coronal mass at 2R$_\odot$   & 2.1 \\
N-hemi. coronal mass in 3D      &    1.9    & N-hemi. coronal mass at 2R$_\odot$  & 2.1 \\
S-hemi. coronal mass in 3D      &    2.0    & S-hemi. coronal mass at 2R$_\odot$  & 2.2 \\
N-pole coronal mass in 3D       &    4.3    & Global streamer area at 2R$_\odot$  &  1.6 \\       
S-pole coronal mass in 3D       &    3.5    & N-hemi. streamer area at 2R$_\odot$ &  1.6 \\
N-royal coronal mass in 3D      &    3.4    & S-hemi. streamer area at 2R$_\odot$ &  1.7 \\
S-royal coronal mass in 3D      &    3.9    & Global streamer mean density at 2R$_\odot$  & 1.4\\
Equator coronal mass in 3D      &    1.6    & N-hemi. streamer mean density at 2R$_\odot$ & 1.5\\
                                &           & S-hemi. streamer mean density at 2R$_\odot$ & 1.5\\
\hline
\end{tabular}
\end{table}

Our result agrees with the study of \citet{don07} who by a wavelet analysis found that the two hemispheres never shifted out of phase by more than $\pm$10 months (or 10\% of the cycle period) over the past 130 years. Historical sunspot records showed that the northern hemisphere has been leading since about 1965 (the start of Cycle 20), and this hemispheric phase-leading appears to be a secular variation with only several changes occurring during the last 300 years \citep{zol10, mci13, hat15}. Several recent studies suggested that the north-south asymmetry of magnetic activity and the persistent one-hemisphere leading the other may be related to the asymmetry of the meridional flow \citep{mci13, zha13, vir14, bla17}. Some nonlinear dynamo models also showed that strong hemispheric asymmetry can be produced by stochastic fluctuations in the dynamo governing parameters \citep[\eg][]{min02, min04, uso09}, or via nonlinear parity modulation \citep[\eg][]{sok94, bee98}. 

The modulation factors are often used to characterize the variability of the coronal radiance over solar cycles. Many previous studies determined these factors based on the global K+F corona, the K-corona, or pB observations and found the typical values in the range of 2--4 (see Table~1 in \opencite{bar15}). The modulation factors vary with the strength of the cycle but also depend on the way the data are averaged. Using 14-CR ($\approx$13 months) running averages, we measure MFs from the increase of the total mass (or average electron density) of the corona during the period from minimum to maximum of Solar Cycle 24 , and obtain MF=1.6--4.3. This measurement agrees well with \citet{bar15}, who obtained MF$_3$=1.5--4.2 from the 13-month running average of K-coronal radiance for the same activity period. We find that the modulation factors are latitude-dependent with the largest in the polar regions. This result also agrees with that of \citet[][see Table 3 in their paper]{bar15}. Note that using COR1 data with the {\scshape /calroll} background subtraction may lead to an overestimate of the MF, in particular in the polar region due to underestimation of the radiance as discussed above. However, this effect appears to be trivial as our measured MFs are comparable to those from LASCO/C2. We also find the modulation factors show a hemispheric asymmetry: MFs in the northern hemisphere and northern royal zone are smaller than in the southern hemisphere and its southern royal zone, but the MF at the North Pole is larger than that at the South Pole. In addition, we measure the variation of streamers, and find that the modulation factors of their total mass depend on the changes in both their total area and average density.

We analyze the short-term fluctuations of the coronal mass (or coronal electron density) during the rising and maximum epochs of Cycle 24, and determine the oscillation periods to be 8--9 CRs (7--8 months) using wavelet analysis. The oscillations of the streamer total mass appear to be mainly determined by their mean density oscillations. Our measured periodicities are consistent with those obtained by \citet{bar15} from LASCO/C2 data. \citet{bar15} also found that the oscillation periods over Cycle 23 are about one year, and that these quasi-periodic oscillations are highly correlated with those of the photospheric total magnetic flux. Multiple periodicities of solar activity, characterized with variable quasi-periods in the range of 0.6--4 years present in all levels of the solar atmosphere, are known as quasi-biennial oscillations (QBOs) (see a comprehensive review by \inlinecite{baz14}). These QBOs are probably linked through the magnetic field, and their origin and periodicities may be associated with stochastic processes of (active region) magnetic flux emergence during the solar cycle \citep[\eg][]{rie84, wan03, hat15}.

We determine the radial electron density distributions of streamers at solar minimum (from 2007/12 to 2009/10) and maximum (from 2011/8 to 2014/7) of Cycle 24, and find that the average density at solar maximum is only slightly larger (by $\approx$30\%) than that at solar minimum. This result was not due to the choice of calculating average densities over the areas with $N_{\rm e}>3\sigma$. The averages for the streamer regions with $N_{\rm e}>1\sigma$ or 6$\sigma$ give a similar result. By comparison with some previous electron density models of solar minimum such as the \citet{sai77} model ($N_{\rm Saito}(r)$) and the \citet{guh96} model ($N_{\rm GHM}(r)$) based on observations (1973/5--1974/2) during the declining phase of Cycle 20 near solar minimum, the \citet{leb98} model ($N_{\rm Leblanc}(r)$) in the period of 1994/12--1997/11 near the minimum of Cycle 22, and the \citet{gib99} model ($N_{\rm Gibson}(r)$) in 1996/8, our derived solar minimum electron density model ($N(r)_{\rm min}$) is lowest in value (having the average ratios $N_{\rm Leblanc}/N_{\rm min}\approx$1.3, $N_{\rm Gibson}/N_{\rm min}\approx$1.8, and $N_{\rm Saito}/N_{\rm min}\approx$2.2 over 1.5--3.0 $R_{\odot}$). This can be explained by the fact that this recent solar minimum was observed with a very low solar activity \citep{mci13, bis14}. \citet{lam14} showed that the global radiance of the K corona was 24\% fainter during minimum of Cycle 23/24 than during that of Cycle 22/23. \citet{dep15} found that the equatorial coronal electron densities obtained using both tomography and thermodynamic MHD model were lower during 2008--2010 than during 1996--1998. Both these studies support our suggestion. In addition, the significant difference of radial density between different models (\eg~$N_{\rm GHM}/N_{\rm min}\approx$3.6) also partially arises from the fact that different features of coronal structures are analyzed.  

In conclusion, we study the long-term and short-term variations of the global K-corona activity in terms of the total coronal mass or mean electron density for Solar Cycle 24. We find a hemispheric asymmetry in both phase and strength: the northern hemisphere leading the southern hemisphere by a shift of 7--9 months, although the former appears to be weaker than the latter as indicated by the modulation factors. The corona shows a conspicuous quasi-periodicity of 7--8 months during the rising and maximum times. The radial distribution of mean electron density for streamers at this solar maximum is only slightly larger than at the minimum.  

\appendix
{\bf The STEREO/COR1 Background Subtraction and its Effect on the Coronal Mass Estimates}

\begin{figure}
\centerline{\includegraphics[width=0.9\textwidth,clip=]{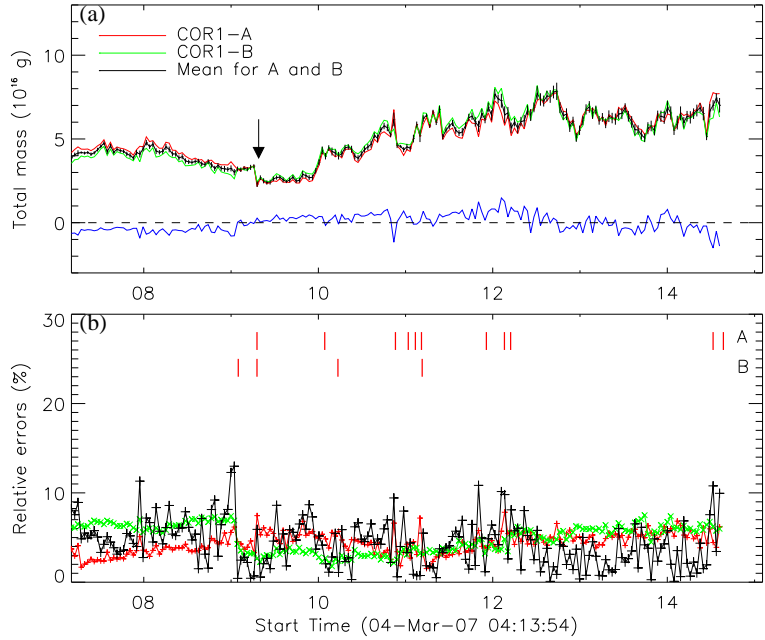}}
\caption{ Estimate of uncertainties in the measured total coronal mass from STEREO/COR1 in the case when the pB images are processed with the regular monthly minimum background images. The annotation for all curves is same as Figure~\ref{fig:err}. The red bars at the top of panel (b) indicate the times of the events that caused a significant change of the scattered light background for COR1-A (upper row) and COR1-B (lower row).} 
\label{fig:reg}
\end{figure}

Removing instrumental stray light (referred to as background subtraction) is an essential step in the pB data reduction because raw COR1 signals are comprised of three components: the K coronal light, the scattered light (weakly polarized), and the F coronal light (unpolarized). The procedures for deriving the time-dependent COR1 instrumental background were described in detail by \citet{tho10}. Using their methods two types of background images (namely the monthly minimum backgrounds and the combined monthly minimum and calibration roll backgrounds) are generated every 10 days for each of the polarizer settings at 0$^{\circ}$, 120$^{\circ}$, and 240$^{\circ}$. In SSW the routine {\scshape secchi\_prep} is used to calibrate the COR1 data, including the process of background subtraction with a choice of using the regular monthly minimum backgrounds by default or using the combined background with the keyword {\scshape /calroll}.  Table~\ref{tab:calr} lists the dates for the COR1 calibration roll maneuver during the period of our interest. The comparison of Figure~\ref{fig:err} with Figure~\ref{fig:reg} shows that the differences of the total coronal mass calculated for COR1-A and -B are relatively larger due to use of the calibration roll backgrounds compared to the case using the regular monthly minimum backgrounds. This may be because calibration rolls of COR1-A and -B were performed typically four times a year and were also out of phase, and the calibration roll background images at other times than those listed in Table~\ref{tab:calr} have to be derived by interpolations (or extrapolations if a background change event occurred between the two closest calibration roll maneuvers). Long-term monitoring reveals that the COR1 background occasionally encounters a sudden increase, which is most likely due to a dust particle landing on the objective lens. For example, the dust landing event on 19 April 2009 for COR1-B is the biggest one, and other events that significantly affected the COR1 background are listed in Table~\ref{tab:evt}, where those due to the changes of image binning format and exposure time are also included. However, it needs to be cautioned that the background subtraction close to these events is generally poorer than normal, which may lead to the relatively larger uncertainties in the total coronal mass estimated around these events (see Figure~\ref{fig:err}b and Figure~\ref{fig:reg}b).

\begin{table}
\caption{ The dates for COR1 calibration rolls performed during the period when our analyzed data are observed.}
\label{tab:calr}
\begin{tabular}{llll}
\hline
  COR1-A    &   COR1-B               &   COR1-A    &         COR1-B \\
\hline
                  &  2007-04-17  &  2011-04-05     &       2011-05-03\\
2008-01-03        &  2008-02-19  &  2011-05-03     &       2011-07-26\\
2008-04-01        &  2008-05-20  &  2011-07-26     &       2011-11-08\\
2008-06-26        &  2008-06-25  &  2011-11-29     &                 \\  
2008-09-30        &  2008-08-26  &  2012-01-10     &       2012-02-14\\ 
2008-12-02        &  2008-12-16  &  2012-04-03     &       2012-05-29\\
2009-03-10        &  2009-02-17  &  2012-06-26     &       2012-09-04\\
2009-06-09        &  2009-04-07  &  2012-09-18     &       2012-12-04\\
2009-09-10        &  2009-06-16  &  2012-12-18     &                \\
2009-11-24        &  2009-09-30  &  2013-03-19     &      2013-03-12\\ 
2010-02-23        &  2010-01-19  &  2013-06-11     &      2013-06-18\\
2010-05-18        &  2010-04-06  &  2013-09-03     &      2013-09-17\\ 
2010-08-10        &  2010-08-03  &  2013-12-13     &      2013-12-26\\
2010-11-09        &  2010-10-08  &  2014-02-11     &      2014-03-25\\
2010-12-16        &              &  2014-05-20     &      2014-07-15\\
2011-02-01        &  2011-02-08  &  2014-08-12     &                \\
\hline
\end{tabular}
\end{table}

\begin{table}
\caption{ Events that significantly change the COR1 background during the period from 2007/03/04 to 2014/08/07.  }
\label{tab:evt}
\begin{tabular}{llll}
\hline
 Time  (UT)               &   COR1-A events          &  Time  (UT)          &  COR1-B events \\
\hline
                    &                          & 2009-01-30 16:20   & dust particle landing   \\
2009-04-19 00:00  & change to 512$\times$512 & 2009-04-19 00:00   & change to 512$\times$512 \\
2010-01-27 16:49  & dust particle landing    & 2010-03-24 01:17   & dust particle landing   \\
2010-11-19 16:00  & dust particle landing    &                      &          \\
2011-01-12 12:23  & dust particle landing    &                      &          \\
2011-02-11 04:23  & dust particle landing    &                      &          \\
2011-03-08 17:00  & dust particle landing    &  2011-03-11 18:50  & dust particle landing \\
2011-12-05 12:03  & dust particle landing    &                      &          \\
2012-02-19 02:33  & dust particle landing    &                      &          \\
2012-03-16 00:00  & exposure time changing   &                      &          \\
2014-07-11 16:00  & dust particle landing    &                      &          \\
2014-08-23 17:00  & dust particle landing    &                      &          \\
\hline
\end{tabular}
\end{table}

\begin{figure}
\centerline{\includegraphics[width=0.9\textwidth,clip=]{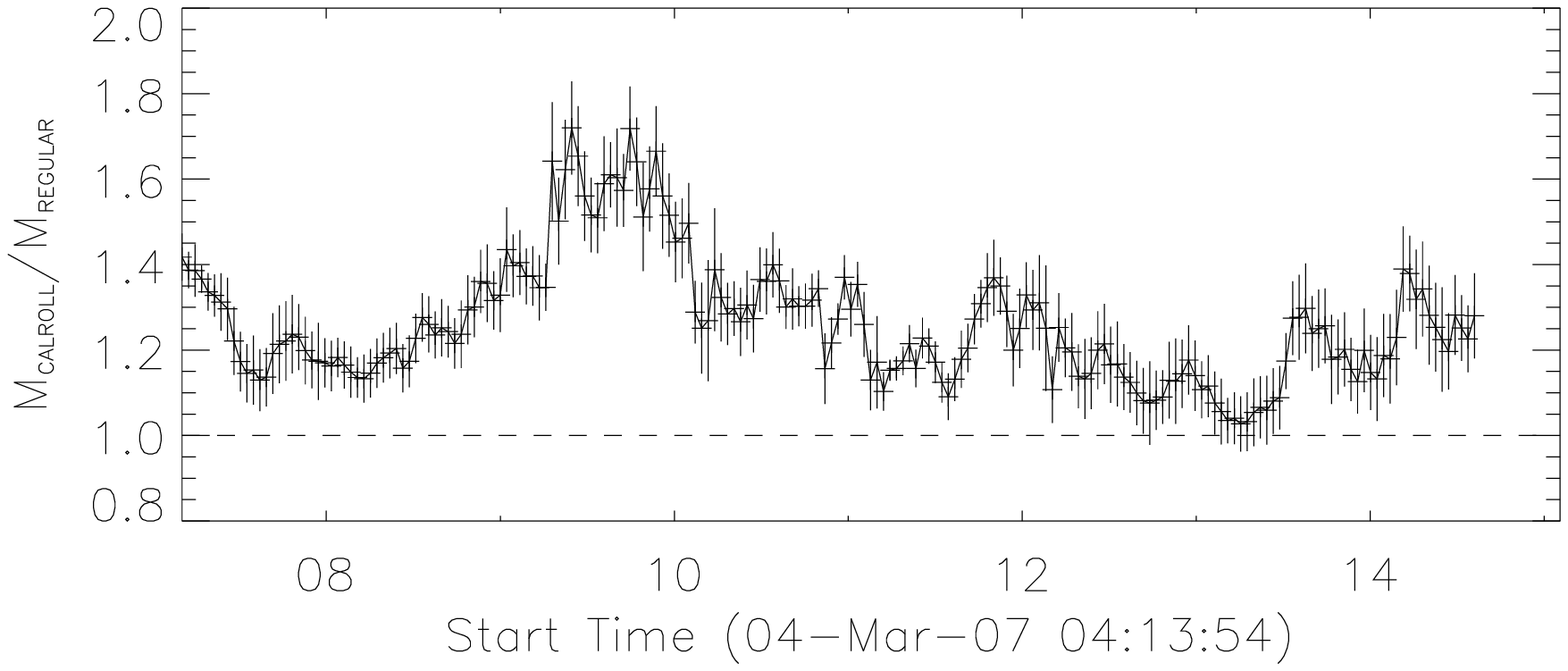}}
\caption{ Ratios of the total coronal mass ($M_{\rm CALROLL}$) measured from the COR1 pB images which are processed with the combined monthly minimum and calibration roll backgrounds to that ($M_{\rm REGULAR}$) measured from the COR1 pB images which are processed with the regular monthly minimum backgrounds. The error bars are calculated using error propagation rules.} 
\label{fig:rat}
\end{figure}
%
\begin{acks}
The work of TW and NLR was supported by the NASA Cooperative Agreement NNG11PL10A to CUA. LASCO was built by a consortium of the Naval Research Laboratory (USA), the Max-Planck-Institut f\"{u}r Sonnensystemforschung (Germany), the Laboratoire d’Astrophysique de Marseille (France), and the University of Birmingham (UK). SOHO is a project of joint collaboration by ESA and NASA. The sunspot data used in this paper were obtained from the World Data Center SILSO, Royal Observatory of Belgium, Brussels. We are grateful to the anonymous referee for the constructive comments that improved the manuscript.
\end{acks}

\vspace{3mm}
\noindent
{\bf Disclosure of Potential Conflicts of Interest}
The authors declare that they have no conflicts of interest.


\end{article} 
\end{document}